\documentclass[useAMS,usenatbib,referee]{mn2e}
\usepackage{epsfig}
\usepackage{graphicx}
\usepackage{lscape}
\usepackage{color}

\title[ Population study   for  $\gamma$-ray pulsars]
{ Population study for  $\gamma$-ray pulsars: (III)  Radiation
 characteristics and  viewing geometry}
\author[J. Takata, Y. Wang and K.S. Cheng]{J. Takata \thanks{E-mail:
takata@hku.hk}, Y. Wang \thanks{E-mail:yuwang@hku.hk}  and  K.S. Cheng\thanks{E-mail:hrspksc@hkucc.hku.hk}\\
Department of Physics, University of Hong Kong,
Pokfulam Road, Hong Kong
} 
\begin{document}

\date{}

\pagerange{\pageref{firstpage}--\pageref{lastpage}} \pubyear{2010}

\maketitle

\label{firstpage}

\begin{abstract}
 We have performed a Monte-Calro simulation for  Galactic population 
of pulsars and  for  the $\gamma$-ray observations. 
 We apply  two-layer outer gap model, which has been developed 
by Wang, Takata \& Cheng,  for 
the $\gamma$-ray emission process, and study the  radiation characteristics 
  as a function of  the magnetic inclination angle $(\alpha)$
 and the Earth viewing angle ($\zeta$). In our model,  
the $\gamma$-ray flux and the spectral 
cut-off energy tend to decrease as the inclination and 
 viewing angles deviate from $90^{\circ}$. 
The emerging spectrum above 100~MeV becomes soft with 
a photon index  $p\sim 1.8-2$ for $\zeta \rightarrow 90^{\circ}$ and 
$p\sim 1.2-1.3$ for $\zeta \ll 90^{\circ}$. 
Our simulation  predicts  that  
the pulsars with larger inclination angles ($\alpha=70-90^{\circ}$) and 
larger viewing angles ($\zeta=70-90^{\circ}$) have been preferentially detected 
by the $Fermi$ $\gamma$-ray telescope, 
and hence  the observed pulse profiles of the 
 $\gamma$-ray pulsars have  the double peak structure rather 
than  single peak.   In the simulation, 
most  $\gamma$-ray millisecond 
pulsars are categorized as the radio-quiet $\gamma$-ray pulsars, because 
its radio fluxes are under the sensitivities of  the 
major radio surveys. Even we drastically increase the radio 
sensitivity by a factor of ten, the number of radio-selected 
millisecond pulsars detected by the $Fermi$  ten~years observations is still 
much less than the expected  $\gamma$-ray-selected millisecond pulsars,
 indicating  the 
radio-quiet millisecond pulsars must contribute to the $Fermi$ unidentified 
sources and/or the $\gamma$-ray background radiations. 
We argue that  $\gamma$-ray pulsars observed with 
a smaller viewing angle ($\zeta\ll 90^{\circ}$) 
will appear as  low-efficient $\gamma$-ray 
pulsars.  For example,  unique radiation properties of 
the low-efficient $\gamma$-ray pulsar, PSR~J0659+1414,  
can be explained by the present our gap model with a viewing geometry of   
$\alpha\sim \zeta=40^{\circ}-50^{\circ}$.

\end{abstract}


\section{Introduction}
\label{intro}
The Large Area Telescope ($LAT$) on board the $Fermi$ $\gamma$-ray telescope 
has increased  number of $\gamma$-ray pulsars, and the recent $Fermi$ catalog includes more that  60 $\gamma$-ray pulsars, which includes 
9 millisecond pulsars (Abdo et al. 2009a,b, 2010a\footnote{see also 
$\mathrm{http://fermi.gsfc.nasa.gov/ssc/data/access/lat/1yr_{-}catalog}$}; Saz~Parkinson et al. 2010).
  Furthermore, the detection of  radio millisecond pulsars associated with 
 about 20 unidentified $Fermi$ point sources (e.g. Ray 2010; Caraveo 2010; Ranson et al. 2011; Keith et al. 2011) has been reported, 
suggesting  that the millisecond pulsar,  as well as the canonical pulsar,   
is  one of the major Galactic $\gamma$-ray source.  It can be expected  that 
more $\gamma$-ray pulsars will be added to the list over the  $Fermi$ mission.
  The spectral shape and the pulse morphology    measured by the 
 $Fermi$   have  been used to 
discriminate particle acceleration and  $\gamma$-ray emission models; the 
polar cap model (Ruderman \& Sutherland  1975; Daugherty \& Harding 1982, 1996),
 the slot gap model (Arons 1983; Muslimov \& Harding 2004; Harding et al. 2008;
  Harding \& Muslimov 2011) and the outer gap model 
(Cheng, Ho \& Ruderman 1986a,b; Hirotani 2008; Takata, Wang \& Cheng 2010a).
 The polar cap model assumes the acceleration region near the stellar surface, 
and the slot gap and  outer gap models  assume the emission region extending 
to  outer magnetosphere.  The cut-off features of the $\gamma$-ray spectra 
of the Crab and the Vela pulsars measured 
by $Fermi$  imply that the $\gamma$-ray 
emission site of the canonical pulsars
 is located in the outer magnetosphere rather than 
near polar cap region, which  produces  a cut-off feature  
steeper than the observed one (Aliu et al. 2009; Abdo et al. 2009c, 2010d). 
Romani \& Watters (2010) and Watters \& Romani (2011) have studied  
 morphology of  the pulse profiles of the young pulsars  predicted 
by the our gap and slot gap models,  and they  argued statistically 
that the outer gap geometry is  more
  consistent with the $Fermi$ observations than the slot gap model. 
  Venter, Harding \& Guillemot (2009) 
found that the observed pulse profiles of several  millisecond pulsars
detected by the $Fermi$  cannot be  explained by the outer gap and/or  the slot gap  models, and 
 proposed a pair-starved polar cap model, in which the particles are
 continuously accelerated up to high altitude because of the insufficient
 multiplicity of the pairs. 

The high quality data measured by the $Fermi$ 
enables  us to perform a  detail study 
for   population of the $\gamma$-ray pulsars.  
Takata et al.  (2010a) have studied the relation between 
the emission properties (luminosity and spectral cut-off energy) and pulsar 
characteristics (e.g. rotation period and magnetic field). They proposed 
 that the outer gap accelerator model controlled by 
the magnetic pair-creation process  can explain  the observed population 
statistics better than that  controlled by  the photon-photon 
pair-creation process (Zhang \& Cheng 1997, 2003).
 Wang, Takata  \&  Cheng (2010) have fitted  the observed phase-averaged 
spectra by using  a  two-layer  outer gap model, in which the accelerator 
consists of a wide but low charge density main region and a narrow 
 but high charge density 
screening region. They suggested  that the relation between  the $\gamma$-ray 
luminosity ($L_{\gamma}$) and the spin down power ($L_{sd}$) 
can be expressed as $L_{\gamma}\propto L_{sd}^{\beta}$ with $\beta\sim 0$ for 
$L_{sd}\ge  10^{36}$~erg/s, while $\beta\sim 0.5$
 for $L_{sd}\le  10^{36}$~erg/s. This relation is consistent with 
the theoretical expectation, i.e. the gap fractional size $f$ 
is determined by  $f=min(f_{zc}, f_m)$, where  $f_{zc}$ and $f_{m}$ 
are  the gap size determined by  photon-photon pair-creation process  
(Zhang \& Cheng 1997, 2003) and  magnetic pair-creation  process 
(Takata et al 2010a), respectively (c.f. section~\ref{geometry}).
 Equating $f_{zc}$ and $f_m$ corresponds to $L_{sd}\sim 10^{36}$~erg/s. 

The study for population synthesis 
 of the radio  pulsars has been developed by using the detailed modeling
 of the radio emissions  and the radio surveys
(e.g. Bailes \& Kniffen 1992; Sturner and Demer 1996;
 Faucher-Gigu$\grave{\mathrm{e}}$re and  Kaspi, 2006).   
Also the population study of the $\gamma$-ray pulsars have  
been developed by several authors (e.g. 
  Gonthier et al. (2002) for the polar cap model,  Cheng \& Zhang 
(1998) for the outer gap model, and Story, Gonthier \& Harding (2007) 
for the slot gap model).  For example, 
Story et al. (2007) studied the 
population of $\gamma$-ray millisecond pulsars with
the slot gap accelerator model, and predicted the $Fermi$ observations.
 They predicted that the $Fermi$ will detect 12 radio-loud and 
33-40 radio-quiet $\gamma$-ray millisecond pulsars.  
With the Monte-Calro simulation of the outer gap,  
Takata, Wang \&  Cheng (2011a,b) have explained 
the observed  distributions of the characteristics 
of the  $\gamma$-ray pulsars detected by the $Fermi$ with the 
six-month long  observations. They predicted that at least 80 $\gamma$-ray 
canonical  pulsars have a $\gamma$-ray flux exceeding 
 the sensitivity of the six-month $Fermi$ observations, suggesting  
 the present observations have missed  many $\gamma$-ray emitting pulsars. 
 Watters \& Romani (2011) simulated 
the Galactic distribution of the young $\gamma$-ray pulsars, and compared 
the simulated pulse morphology   with the $Fermi$ results. 
They argued statistically that the outer gap model explains  
the distributions of 
the pulse morphology  (e.g. the phase-separation of the two peaks)
 measured by the $Fermi$ better than the slot gap model.  
The  population studies  (e.g.	Kaaret \&  Philip 1996;
 Cottam, Jean  Faucher-Gigu$\grave{\mathrm{e}}$re 
\&  Loeb 2010; Takata et al. 2011b)  have also pointed 
out that unidentified pulsars, in particular the millisecond pulsars located 
at high-Galactic latitudes will associate with the $Fermi$ 
unidentified sources (Abdo et al. 2010b),
 and will  contribute to  the $\gamma$-ray background radiations.

In the our previous studies (Takata et al. 2011a,b), we 
ignored  the dependence of the radiation characteristics 
on the viewing geometry (i.e. the  magnetic inclination angle and 
the Earth viewing angle measured from the pulsar's rotation axis), 
and focused on the  distributions of the 
properties  (e.g. the  rotation period and  magnetic field) 
of the $\gamma$-ray pulsars.   However,  
the observed radiation characteristics,  such as the flux, spectral  
cut-off energy  and  pulse profile,  
  must be affected by  the viewing geometry. To perform  a more solid  study on 
the population, which is compared with the high-quality  $Fermi$ data,  
it  is required a three-dimensional model that takes into account
 the dependence of the radiation characteristics on the viewing geometry. 

In this paper,  we develop a Monte-Carlo study for the population 
of the $\gamma$-ray pulsars with the radiation model that takes into account  
 the dependence on the viewing geometry. 
In section~\ref{simulation}, we review the Monte-Carlo simulation for 
the population of the $\gamma$-ray pulsars. We discuss our outer gap model 
in  section~\ref{gemission}.  
In section~\ref{dependency}, we discuss  the 
 dependence of the radiation characteristics on the viewing geometry. 
In section~\ref{monte}, we compare the results of the Monte-Calro simulation 
with the  $Fermi$ six-month long observations.
 We also show 
the expected population of the $\gamma$-ray pulsars if the $Fermi$ 
observations continue  five-years or ten-years.  In section~\ref{summary}, 
after summarizing our simulation results, 
we discuss the viewing geometry of PSR~J0659+1414, which is 
known as a  low efficient $\gamma$-ray pulsar.

\section{Monte-Carlo simulation for the pulsar population}
\label{simulation}
In this paper, we denote the canonical pulsar and the millisecond pulsar 
as CP and MSP, respectively. 
We assume that  the birth rates of the CPs and the MSPs
 are  $\sim 10^{-2}\mathrm{yr^{-1}}$ and 
 $10^{-6}\sim 10^{-5}~\mathrm{yr^{-1}}$ (Lorimer et al. 1995, Lorimer 2008), 
respectively, and we ignore the millisecond pulsars in globular clusters. 
 The birth location 
 is determined  by the  spatial distributions given by  
(Paczynski 1990), 
\[
\rho_R(R)=\frac{a_{R}\mathrm{e}^{-R/R_{\mathrm{exp}}}R}{R^2_{\mathrm{exp}}},
\]
\begin{equation}
\rho_Z(Z)=\frac{1}{Z_{\mathrm{exp}}}\mathrm{e}^{-|Z|/Z_{\mathrm{exp}}},
\end{equation}
where $R$ is the axial distance from the axis through the Galactic
centre  perpendicular to the Galactic disk and  $Z$ is the distance from
the Galactic disk,   
$R_{\mathrm{exp}}=4.5$~kpc and 
$a_R=[1-\mathrm{e}^{-R_{max}/R_{exp}}(1+R_{max}/R_{exp})]^{-1}$ with $R_{max}=20$
~kpc. In addition, we apply  $Z_{exp}=75$~pc for the CPs 
and $Z_{exp}=200$~pc for the MSPs, respectively.

To obtain  current position of each simulated pulsar, we 
solve the  equation of motion from its birth to the current time. 
The equation of motion is given by  
\begin{equation}
 \frac{dR^2}{dt^2}=\frac{v_{\phi}^2}{R}-\frac{\partial
 \Phi_{tot}}{\partial R},
\label{eqr}
\end{equation}

\begin{equation}
 \frac{dZ^2}{dt^2}=-\frac{\partial \Phi_{tot}}{\partial Z},
\label{eqz}
\end{equation}
and 
\begin{equation}
Rv_{\phi}=\mathrm{constant}.
\end{equation}
Here $v_{\phi}$ is the azimuthal component of the velocity,  $\Phi_{tot}=\Phi_{sph}+\Phi_{dis}+\Phi_h$ is the total gravitational potential, where 
$\Phi_{sph}$, $\Phi_{dis}$ and $\Phi_{h}$ are spheroidal, disk
and halo components of the Galactic gravitational potential, and  are
given  by
\begin{equation}
\Phi_i(R,Z)=-\frac{GM_i}{\sqrt{R^2+[a_i+(Z^2+b_i^2)^{1/2}]^2}},
\end{equation}
where $i=sph$ and $dis$,  $a_{sph}=0$, $b_{sph}=0.277$~kpc, 
$M_{sph}=1.12\times 10^{10}M_{\odot}$, $a_{dis}=3.7$~kpc, $b_{dis}=0.20$~kpc,
and $M_{dis}=8.07\times 10^{10}M_{\odot}$,  while for the halo 
component 
\begin{equation}
\Phi_{h}(r)=-\frac{GM_{c}}{r_c}\left[\frac{1}{2}\ln
 \left(1+\frac{r^2}{r_c^2}\right)+\frac{r_c}{r}\tan^{-1}
\left(\frac{r}{r_c}\right)\right],
\end{equation}
where $r_c=6.0$~kpc and $M_c=5.0\times 10^{10}M_{\odot}$ (c.f. Burton \& Gordon 1978; Binney \& Tremaine 1987; Paczynski 1990). 
The Lagrangian in units of energy
per unit mass is given by
\begin{equation}
L=\frac{v^2(R,Z,\phi)}{2}-\Phi_{tot}(R,Z),
\end{equation}
where $v$ is the velocity.

For the initial velocity of each modeled pulsar, 
we assume  random isotropic direction of the velocity, 
 and  assume magnitude drawn from 
a  Maxwellian 
distribution with a characteristic width of  
$\sigma_V=265$~km/s for the CP and $\sigma_V=70$~km/s for 
the MSP (c.f.  Hobbs et al. 2005), namely,
\begin{equation}
\rho_V(V)=\sqrt{\frac{\pi}{2}}\frac{V^2}{\sigma_V^3}
\mathrm{e}^{-V^2/2\sigma_V^2}.
\end{equation}

\subsection{Pulsar characteristics}
In this section, we describe how we calculate  the current value of the various 
characteristics of the simulated pulsars. 
\subsubsection{Canonical pulsars}
For canonical $\gamma$-ray pulsar, 
which is born soon after the supernova explosion,
 the true age   is more or less equal to the spin down age $\tau=P/2\dot{P}$, 
where $P$ and $\dot{P}$ are the rotation period and its time derivative.   
This allows us to calculate the present distributions of the pulsar 
characteristics from the initial distributions. 

For the Crab pulsar, the initial period is estimated to be 
 $P_0\sim 19$ms. A short birth rotation period  is also 
expected for the  young 16ms pulsar PSR~J0537-0691 (Marchall et al. 1998), 
suggesting most pulsars were  born with $P_0\sim 20-30$~ms.
There is a  good evidence 
 that  some pulsars were born with a longer initial period (e.g. 
$P_0\sim$62~ms of PSR~J1911-1925, Kaspi et al. 2001). 
{ However, we found  that the distribution of initial rotation 
 period does not affect 
much to the  results of the following Monte Carlo simulation.
 For example, we compared the results of two simulations with different 
 distribution of  the initial period; one is that all simulated pulsars are 
randomly distributed in the range  $P_0=20-30$~ms, and other is that 
 70~\% of pulsars are distributed at $P_0=20-30~$ms and 
 30~\% are  $P_0=20-100$~ms.  In such a case, we could not see any significant  
difference  in the  populations of the simulated $\gamma$-ray pulsars.  
The difference in  the simulated distributions of  the $\gamma$-ray pulsars 
 becomes significant if we assume that $>50~\%$ of new born pulsars 
is  distributed with the initial period of  $P_0\ge 30$~ms. 
Because  we do not know well the exact 
distribution of the initial period, we randomly choose 
the initial period in the narrow range $P_0=20-30$~ms, 
which provides  a more consistent  result of our simulation 
 with the $Fermi$  observations. 

We assume a Gaussian distribution in $\mathrm{log}_{10}B_s$ for the initial
 distribution of the stellar  magnetic field measured at the magnetic equator, 
\begin{equation}
\rho_B(\mathrm{log}_{10}B_s)=\frac{1}{\sqrt{2\pi}\sigma_B}
\exp\left[-\frac{1}{2}\left(\frac{\mathrm{log}_{10}B_s-\mathrm{log}_{10}B_0}
{\sigma_B}\right)^2\right].
\end{equation}
In this study, we apply $\mathrm{log}_{10}B_0=12.6$ and $\sigma_B=0.1$. 
Because the canonical $\gamma$-ray pulsars are younger than 10~Myr, we ignore 
 evolution of the stellar magnetic field, which may be important 
for the neutron star with an age older 
 than 10~Myr (Goldreich \& Reisenegger 1992;  Hoyos, Reisenegger \& Valdivia,  2008). 

With a constant stellar magnetic field in time, the period evolves as 
\begin{equation}
P(t)=\left(P_0^2+\frac{16\pi^2R_s^6B_s^2}{3Ic^3}t\right)^{1/2},
\end{equation}
where $R_s=10^{6}$~cm is the 
stellar surface and $I=10^{45}\mathrm{gcm^2}$ is the neutron star momentum 
of inertial.   In addition, we artificially assumed that the magnetic 
inclination angle does not evolve with the spin down age, 
while a decreases on the spin down time scale has been pointed out 
(e.g. Davis \& Goldstein 1970; Michel 1991; Tauris \& Manchester 1998). 
The time derivative of the rotation period is calculated 
from
\begin{equation}
\dot{P}(t)=\frac{8\pi^2R_s^6B_s^2}{3Ic^3P}.
\end{equation}

\subsubsection{Millisecond pulsar}
We assume that all MSPs are born through the so called recycled process, 
in which the accretion of the matter from the low mass companion star 
spins up the neutron star. It implies that the true age of the binary system 
is different from the spin down age of the MSP. 
However, we  expect that the Galactic distribution does  not depend 
on the spin down age of the MSPs.  With the typical velocity of the
observed MSPs ,  $V\sim 70$~km/s,  it is expected that the 
displacement of the MSPs (or binary system) 
with the typical  age, $\ge 100$~Myr, 
becomes larger than the size of the Galaxy. With the  relatively 
 slow velocity, $V\le 100$~km/s,  however, 
the MSPs remain bound in the Galactic potential  and hence
their Galactic distribution does  not  depend on the spin down age. 

The initial  rotation period of  MSP is related with the 
history of the accretion process after the decaying stage of the magnetic 
field (Campana et al. 1999; Takata, Cheng and Taam 2010b).
 However, the description of the 
transition from an accretion powered to the rotation powered phase is 
not well understood due to the complexities in the description 
of the interaction between the magnetosphere 
of a neutron star and its accretion disk (Romanova et al. 2009).
Furthermore, the true age of the MSP after the supernova explosion 
in the binary system is different from the spin down age of the MSP.  
These theoretical uncertainties  make it difficult   
 to obtain the present distributions of the MSP   
 properties  from the initial distributions.

In this paper, we assign the ``current''
 pulsar properties   for each
simulated MSP, instead of modeling from the initial distributions; namely, 
 we (1) randomly select the age of the simulated
MSP  up to $10$~Gyr,  (2) shifts the simulated MSP from its birth location 
 to the current location,   and (3) assign the  parameters of the MSP 
following the observed  distributions. 
We assign  the period time derivative ($\dot{P}$)
and the stellar magnetic field ($B_s$) following the observed
$\dot{P}-B_s$ distribution (Manchester et al. 2005).
  From the assigned  period time derivative 
and the stellar magnetic field, the current rotation period and the 
spin down age are calculated from 
\begin{equation}
P_{-3}=0.97B_8^2\dot{P}_{-20}^{-1}~\mathrm{ms}
\end{equation}
and
\begin{equation}
\tau=1.5\times 10^9 P_{-3}\dot{P}_{20}^{-1}~\mathrm{yr}
\label{age}
\end{equation}
respectively (Lyne \& Graham-Smith, 2006).
Here $B_8$ is the stellar magnetic field in units of $10^8$~G, 
 $P_{-3}$ and $\dot{P}_{-20}$ are the rotation period in units
of 1~millisecond and its  time derivative in units of $10^{-20}$, respectively.
 In fact, above process  can provide 
a consistent  Galactic distribution of the radio MSPs with the 
observations (c.f. Takata et al. 2011b).   
\subsection{Radio emissions}
Using  the empirical relation 
among the radio luminosity, rotation period, and period time derivative, 
 the distribution of the
 radio luminosity  at  400~MHz is expressed by (Narayan \& Ostriker 1990) 
\begin{equation}
\rho_{L_{400}}=0.5\lambda^2\mathrm{e}^{\lambda},
\end{equation}
where  $\lambda=3.6[\mathrm{log_{10}}(L_{400}/<L_{400}>)+1.8]$ with
$<L_{400}>=\eta 10^{6.64}\dot{P}^{1/3}/P^3$, and  
$L_{400}$ is the luminosity in units of  $\mathrm{mJy~kpc^2}$.  
Here $\eta$ is a scaling factor to adjust the observed distribution,  
and  $\eta=1$~ (or 0.05) for the CPs (or MSPs). 
 The radio flux on Earth is given by $S_{400}=L_{400}/d^2$, 
where $d$ is the distance to 
the pulsar. We scale the simulated 400~MHz luminosity
 to the observational frequency using 
a typical photon index $\sim 2$ for the CPs and $\sim$1.8 for the MSPs, 
respectively   (Kramer  et al. 1997, 1998).

We take into account the beaming  of the radio emission. 
For the CPs, we apply the half-angle, which is measured from the magnetic axis, 
 of the radio cone studied by Kijak and Gil (1998, 2003), 
\begin{equation}
\omega_{CP}\sim 1^{\circ}.24r_{KG}^{1/2}P^{-1/2},
\label{kg}
\end{equation}
where 
\[
r_{KG}=40\nu_{GHz}^{-0.26}\dot{P}^{0.07}_{-15}P^{0.3}, 
\]
where $\dot{P}_{-15}$ is the period time derivative in units of $10^{-15}$, and 
$\nu_{GH}$ is the radio frequency in units of GHz.  
For the MSPs, the half-angle does not depend on the frequency and 
 is approximately  described as  (Kramer \&  Xilouris 2000),
\begin{equation}
\omega_{MSP} \sim \omega_0(P/1~\mathrm{s})^{-1/2}, 
\label{kx}
\end{equation} 
where $\omega_0$ is randomly chosen in the range  $2.75^{\circ}-5.4^{\circ}$. 
The radio  emission can be detected  by observer with a viewing angle
 between max($0^{\circ}$,$\alpha-\omega_i$)
 and min($\alpha+\omega_i$, $90^{\circ}$), 
where $i=CP$ or $MSP$, and  $\alpha$ is the inclination angle 
between the rotation axis and the magnetic axis.

We use  the ten radio surveys (Molongo 2, Green Band 2 and 3, Arecibo 2 and 3, 
Parkes 1, 2 and MB, Jordell Bank 2 and Swinburne IL), whose
 system characteristics are listed in table~1 of Takata et al (2011a) 
and the references therein.  To calculate the dispersion measure, we apply 
 the Galactic distribution of  electrons studied by  Cordes \& Lazio (2002). 

\section{$\gamma$-ray emission model}
\label{gemission}
The pulsar rotation energy, which is  thought to be essential energy source  
of the $\gamma$-ray emission, can be released by  both the current 
braking torque and
the magnetic dipole radiation. According to the analysis 
of the force-free magnetosphere done by Spitkovsky (2006), 
the spin down power depends
on the inclination angles as $L_{sd}\propto 1+\sin^2\alpha$; in other words,
$L_{sd}$  changes  only by a factor of 2  with the inclination angle.
In the present  Monte-Carlo study, therefore, we ignore the dependence of 
the spin down power on the inclination angle, and apply the conventional 
expression that $L_{sd}=4(2\pi)^4B_s^2R_s^6/6c^3P^4$.

\subsection{Two-layer outer gap model}
\label{oflux}
The outer gap dynamics is controlled by the pair-creation process, which 
produces a charge distribution in the trans-field 
direction (Cheng, Ho \& Ruderman
 1986a,b; Takata, Shibata \&  Hirotani 2004).  
Wang et al. (2010, 2011) argued that the 
 outer gap should be approximately 
 divided into two layers, i.e. the main acceleration region 
starting from the last-open field lines and the screening region lying  at 
the upper part of the gap. In the main acceleration region, 
the charge density is $\sim$10\% 
of the Goldreich-Julian value (Goldreich \& Julian 1969), and 
 a strong electric field  accelerates the electrons and positrons up to 
the Lorentz factor of $\Gamma\sim 10^{7.5}$. 
The accelerated particles  emit  several GeV photons 
via the curvature radiation process. In the screening region,  the large 
number of pairs created by the pair-creation process starts to screen out  
the gap electric field. 
The curvature radiation from the screening pairs produces mainly $\sim$100~MeV 
photons. 

A simple description of the electric field structure inside 
 the two-layer outer  gap is discussed in  Wang et al. (2010, 2011). 
 We denote  $x$, $z$ and $\phi$ 
as the coordinates along the magnetic field line,
perpendicular to the magnetic field line in the poloidal plane and in the 
magnetic azimuth, respectively. We expect that the particle number 
density   increases  exponentially  near the boundary ($z=h_1$) 
between the main acceleration and screening regions (Cheng et al. 1986a,b),
 and that  the charge density is  almost 
constant in the screening region (Hirotani  2006). Hence, we approximately 
describe  the distribution of the charge density  in  the $z$-direction  with 
the step function as follows  
\begin{equation}
\rho(\vec{r})=\left\{
   \begin{array}{ccc}
            \rho_1(x,\phi), &$if$& 0\leq z\leq h_1(x,\phi),\\
            \rho_2(x,\phi), &$if$& h_1(x,\phi)<z\leq h_2(x,\phi),
   \end{array}\right.
\end{equation}
where,$|\rho_1| <|\rho_{GJ}| < |\rho_2|$,  
$z=0$ and $z=h_2$ correspond to  the last-open field line and 
the upper boundary of the gap, respectively. For simplicity,  
we  define the boundary  between main acceleration region and the screening 
region, i.e. $h_1$, with a magnetic  field line 
with such approximation that  $h_1/h_2$ is  constant along the
magnetic field line. The present model 
predicts that the charge density in the screening region should be proportional 
to the Goldreich-Julian charge density (Wang et al. 2010).  This situation will 
be satisfied because of a lot of the pairs created by the 
pair-creation process in the screening region. In the main acceleration region,
 because the number  density is much smaller
 than the Goldreich-Julian value,  its 
distribution along the magnetic field line does not important 
for the electric field distribution. Therefore, we 
approximate that $\rho-\rho_{GJ}\sim g(z,\phi)\rho_{GJ}(\vec{r})$ for both 
main acceleration and screening regions, where 
\begin{equation}
g(z,\phi)=\left\{
   \begin{array}{ccc}
            -g_1(\phi), &$if$& 0\leq z\leq h_1(x,\phi),\\
            g_2(\phi), &$if$& h_1(x,\phi)<z\leq h_2(x,\phi).
   \end{array}\right.
\end{equation}
We assume that $g_1>0$ and $g_2>0$ so that $|\rho|<|\rho_{GJ}|$ 
for the main  acceleration  region and $|\rho|>|\rho_{GJ}|$ for 
the screening region.

To  obtain the  typical strength of the electric field in the gap,  
we find the solution of  the Poisson equation 
for each azimuthal angle (c.f. Wang et al. 2010, 2011),
\begin{equation}
 \frac{\partial^2{}}{\partial{z^2}}\Phi'(x,z,\phi)|_{\phi=\mathrm{fixed}}
=-4\pi [\rho(x,z,\phi)-\rho_{GJ}(\vec{r})]_{\phi=\mathrm{fixed}},
\label{Poisson1}
\end{equation}
where $\Phi'$ is the  electric potential of the accelerating field. 
 Here we assumed that the derivative of the Potential field in the $z$-direction 
is much larger than that of $x$-direction, and of  $\phi$-direction.

In this paper, we neglect 
the $z$-dependence of the Goldreich-Julian charge 
density,  and approximate the Goldreich-Julian charge density as 
$\rho_{GJ}(x,\phi)\sim -\Omega B x/2\pi c R_c$ (Cheng et al. 1986a,b), 
where $\Omega$ and $R_c$ are the angular frequency of the pulsar 
 and the curvature radius of the field line, respectively.   
The boundary conditions on the lower ($z=0$) and upper ($z=h_2$) boundaries 
are given by  
\begin{equation}
\Phi'(x,z=0,\phi)=0~\mathrm{and}~\Phi'(x,z=h_2,\phi)=0
\end{equation}
respectively.  Imposing $\Phi'$ and $\partial\Phi'/\partial z$ are continuous 
at the boundary  $z=h_1$, we obtain the solution as 
\begin{equation}
\Phi'(\vec{r})=-\frac{\Omega B x h^2_2(x,\phi)}{cR_c}\left\{
   \begin{array}{ccc}
            -g_1(\phi)z'^2+C_1z', &$for$& 0\leq z'\leq
	     h_1(x,\phi)/h_2(x,\phi)\\
g_{2}(\phi)
(z'^2-1)+D_1(z'-1), &$for$& h_1(x,\phi)/h_2(x,\phi)\leq z'
\leq 1
   \end{array}\right.
\label{potential}
\end{equation}
where
\[
 C_1(x,\phi)=-
\frac{g_1
h_1(h_1-2h_2)
+g_2(h_1-h_2)^2}
{h_2^2},
\]
\[
 D_2(x,\phi)=-\frac{g_1h_1^2
+g_2h_2^2}
{h_2^2}, 
\]
and  $z'\equiv z/h_2(x,\phi)$. The accelerating electric field, 
$E_{||}=-\partial \Phi'/\partial x$, is writted as 
\begin{equation}
E_{||}(\vec{r})\sim\frac{\Omega Bh^2_2(x,\phi)}{cR_s}\left\{
\begin{array}{lcc}
-g_1(\phi)z'^2+C_1(\vec{r})z',~~$for$~~0\le{}z'\le{}h_1(x,\phi)/h_2(x,\phi), \\
g_2(\phi)(z'^2-1)+D_1(\vec{r})(z'-1),~~
$for$~~h_1(x,\phi)/h_2(x,\phi)<z'\le{}1, 
\end{array}
\right.
\label{electric}
\end{equation}
where we used the relations of the dipole field 
that  $\partial(Bh^2_2)/\partial x\sim 0$, 
$\partial z'/\partial x=\partial(z/h_2)/\partial x\sim 0$, 
$\partial(h_1/h_2)/\partial x\sim 0$, and approximated that 
$\partial R_c/\partial x\sim 0$. 

On the upper boundary,  we anticipate that the total potential field   
(co-rotational potential + non co-rotational potential)  in the 
gap is continuously connected
to the co-rotational potential field outside the gap. This screening 
condition is described by 
\begin{equation}
\frac{\partial \Phi'}{\partial z}|_{z=h_2}=-E_{\perp}(x,z=h_2,\phi)=0.
\end{equation}
This condition gives the relation between $(h_1,~h_2)$ and $(g_1,~g_2)$ as 
\begin{equation}
\left(\frac{h_2}{h_1}\right)^2=1+\frac{g_1}{g_2}.
\label{condition}
\end{equation}
In this paper, we do not consider the azimuthal distribution of the 
dimensionless charge density $g_1$ and $g_2$, because we discuss the general 
properties of the $\gamma$-ray emissions. The azimuthal structure will be   
important for explaining  the detailed observed 
properties,  such as  the existing of the third peak in the pulse profile of the Vela 
pulsar measured by the $Fermi$,  and the energy dependence of  the pulse phase of the third peak 
(c.f. Wang et al. 2011).

The typical Lorentz factor of the accelerated particles can be estimated by 
force balance between the electric field and the curvature radiation drag 
force as 
\begin{equation}
\Gamma=\left(\frac{3R_c^2}{2e}E_{||}\right)^{1/4}.
\end{equation}
The spectrum  of the curvature radiation emitted by the individual 
particle is written as 
\begin{equation}
P_c(E_{\gamma},\vec{r})=\frac{\sqrt{3}e^2\Gamma}{hR_{c}}F(\chi),
\end{equation}
where $\chi=E_{\gamma}/E_c$ with $E_c=3hc\Gamma^3/4\pi R_c$ and 
\[
F(\chi)=\chi\int_{\chi}^{\infty}K_{5/3}(\xi)d\xi,
\]
where $K_{5/3}$ is the modified Bessel function of  order 5/3.  
A  $\gamma$-ray spectrum measured on  Earth may be expressed by 
 (e.g. Hirotani 2008)
\begin{equation}
\frac{dF_{\gamma}}{dE_{\gamma}}\sim \frac{1}{d^2}
\sum_{\vec{r}_i}N(\vec{r}_i)P_c(E_{\gamma},\vec{r}_i)R_c(\vec{r}_{i})
 \triangle A_i,
\end{equation}
where $N\sim|\rho_{GJ}|/e$ is the particle number density, $\vec{r}_i$
 represents the radius to the emission point,  from which the emission 
is  measured by the observer,  $\triangle A_i$ is  
the area of the calculation grid  
 in perpendicular to the magnetic field lines. The integrated energy flux 
between 100~MeV and 300~GeV can be calculated from
\begin{equation}
F_{\gamma,100}=\int_{100MeV}^{300GeV}\frac{d F_{\gamma}}{dE_{\gamma}}dE_{\gamma}.
\end{equation}

In the gap, the curvature photons are emitted in the direction of the 
particle motion, which may be described as (Takata, Chang \& Cheng 2007)
\begin{equation}
             \vec{v}=v_p\vec{B}/B + \vec{r}\times\vec{\Omega}, 
\end{equation}
where the first term represents the motion along the magnetic field line, $v_p$
 is calculated from the condition that $|\vec{v}|=c$, and the second term is 
the co-rotation motion.  
The emission direction measured from the rotation axis (i.e.  viewing angle) 
$\zeta$  and the pulse phase  $\psi$  are calculated from (Yadigaroglu 1997)
\begin{equation}
\left\{
   \begin{array}{ccc}
             \cos{\zeta}=\vec{n}\cdot \vec{e}_{\Omega}\\
             \psi=-\psi_n-\vec{r}/R_{lc}\cdot\vec{n},
   \end{array}\right.
\end{equation}
where $\vec{n}=\vec{v}/v$, $\vec{e}_{\Omega}$ is the unit vector in the direction 
of the rotation axis, and $\psi_n$ is the azimuthal angle of the emission 
direction. 
\subsection{Outer gap geometry}
\label{geometry}
In this paper, we adopt a rotating vacuum dipole field as the magnetosphere, 
 and we  assume  that a strong emission region extends between  
the null charge surface 
($\vec{\Omega}\cdot\vec{B}=0$) of the Goldreich-Julian charge density and 
the radial distance $r=R_{lc}$, where $R_{lc}=2\pi/cP$ is the light 
cylinder radius.  The electrodynamic studies  have pointed out 
that the gap current  
can shift the inner boundary toward the stellar surface 
(e.g. Takata et al. 2004).  However,  because it is expected 
that the curvature radiation below the null charge surface 
appears with a emissivity  
 much smaller than that  above the null charge surface (e.g. Hirotani 2006), 
  we ignore its contribution  to the calculation.

We define the fractional gap thickness measured on 
the stellar surface as, 
\begin{equation}
f\equiv \frac{h_2(R_s,\phi)}{r_p(\phi)},
\label{fraction}
\end{equation}
where $r_p$ is the polar cap radius. Note because the electric field $E_{||}$ 
is proportional to $Bh_{2}^2$, it can be found that $E_{||} \propto f^2$.

Zhang \& Cheng (1997, 2003) have argued  a self-consistent outer gap model 
controlled by  the photon-photon pair-creation process between 
the curvature photons and the X-rays from the stellar surface. 
 They estimated the gap fraction  as 
\begin{equation}
f_{zc,~CP}= \frac{h_2(R_s,\phi)}{r_p(\phi)}\sim 
\frac{D_{\perp}(R_{lc})}{R_{lc}}=5.5(P/1\mathrm{s})^{26/21}(B_s/10^{12}\mathrm{G})^{-4/7}
\label{fzccp}
\end{equation}
for the CPs and 
\begin{equation}
f_{zc,~MSP}=7.0\times 10^{-2}(P/1~\mathrm{ms})^{26/21}(B_s/10^8\mathrm{G})^{-4/7}
\delta r_{5}^{7/2},
\label{fzcmp}
\end{equation}
for the MSPs. Here $\delta r_5$ is the distance (in units of $10^{5}$~cm)
 from the stellar surface to the position where the local magnetic 
field is comparable to 
the dipole field, and it will be $\delta r_5\sim 1-10$~cm.

We note that Zhang \& Cheng (1997, 2003) estimated  the gap fraction 
by a completely vacuum  electric field  
$E_{||}=\Omega Bf^2R^2_{lc}/ cR_{c}$.  
With the same gap fraction, 
 the  solution described by  equation~(\ref{electric}) 
gives an electric field at least   a  factor of four smaller 
than that used in Zhang \& Cheng (1997, 2003).  This difference can 
be important for  the typical energy of the curvature radiation ($E_c$), 
because $E_{c}\propto E_{||}^{3/4}$. In other words, 
if we derive  the gap fraction from the pair-creation condition that
 $E_{X}E_{c}=(m_ec^2)^2$, where $E_X$ is the X-ray photon energy, 
 the present model predicts a fractional gap thickness 
larger than that of Zhang \& Cheng (1997, 2003). In this paper, 
reducing  by a factor of four for the electric field  in  the model of 
Zhang \& Cheng (1997, 2003), we apply the gap fraction increased by a factor of 
 $4^{3/7}\sim 1.8$ from those  in equations~(\ref{fzccp}) and~(\ref{fzcmp}).
 
Takata et al. (2010a) proposed that  the outer gap model can be 
controlled  by the magnetic pair-creation process taken place 
near the stellar surface. 
They argued that  half of particles created in the gap or  injected 
 particles into the outer gap at the outer boundary 
return to  the stellar surface, and these returning particles 
emit  $\sim 100$~MeV photons  near the stellar surface.  
A good fraction of  100~MeV photons can make pairs by the magnetic
pair creation process. They argued that  if the magnetic 
field lines near the surface, instead of
nearly perpendicular to the surface,  are bending side-wards  due to the
strong local field, the pairs created in these local magnetic field
lines can have a pitch angle, which is defined by the angle between 
the directions of  the particle's motion and of the field line, 
smaller than 90$^{\circ}$, which results in an 
outgoing flow of pairs, and hence the pairs control the size of the outer gap.
 We also note that the outgoing flow may be 
produced by the Compton scattering.   With this model, 
the fractional gap thickness is estimated as
\begin{equation}
f_m\sim \frac{D_{\perp}(R_s)}{R_p}= 
0.8 K  P^{1/2},
\label{fm}
\end{equation}
where $K\sim B_{m,12}^{-2}s_7$ is the parameter characterizing  the 
local parameters, i.e., $B_{m,12}$ and $s_7$, which 
 are the local magnetic field 
in units of $10^{12}$G and the local curvature radius in units 
of $10^7$cm, respectively. By fitting the  emission characteristic of the 
 $\gamma$-ray pulsars observed by the $Fermi$,
 they estimated as $K\sim 2$ for the CPs and 
  $K\sim 15$ for the MSPs.  
When   the fractional gap thickness $f_{m}$ is smaller  (or larger) 
than  $f_{zc}$, the magnetic  pair-creation (or photon-photon pair-creation) 
process controls  the gap thickness.

We note that the outer gap should only  exist between the last-open 
field lines and the critical magnetic  field lines that have  
  the null charge points  at  the light cylinder.  
Figure~\ref{polar} shows the polar angle, 
which is measured from the magnetic axis, of 
the polar cap rim ($\theta_p$, solid line) and of the foot points of the critical field lines 
($\theta_c$, dashed line) as a function of the magnetic azimuth. The 
maximum gap fraction is defined by $f_{max}=(\theta_p-\theta_c)/\theta_p$,  
and is represented by  the dotted-line in Figure~\ref{polar}. The 
azimuthal angle $\phi=0$ in Figure~\ref{polar} corresponds to the 
plain spanned by the rotation axis and the magnetic axis. In this paper, 
we anticipate that  the azimuthal expansion of the active gap is limited as 
$f\le f_{max}(\phi)$ and $-90^{\circ}\le \phi\le 90^{\circ}$.

\begin{figure}
\begin{center}
\includegraphics{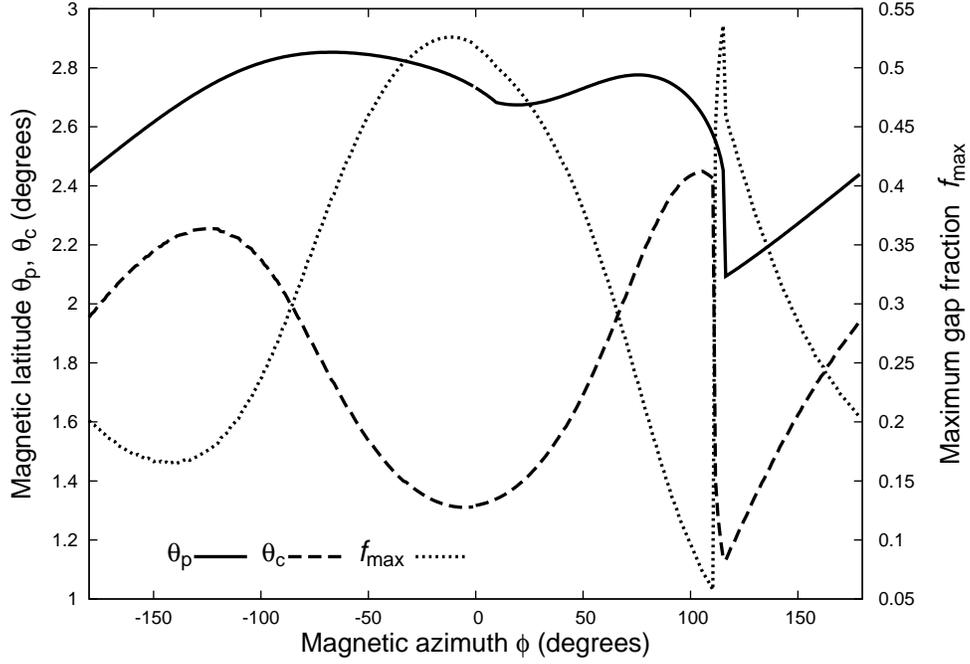}
\caption{Latitudes of the polar cap rim $\theta_p$ (solid line) and of the 
critical magnetic field lines $\theta_c$ (dashed line). The dotted line 
shows the maximum gap fraction $f_{max}=(\theta_p-\theta_c)/\theta_p$. The 
magnetic azimuth $\phi=0$ corresponds to the magnetic meridian, which includes 
the rotation  and magnetic axises. The results are for the inclination angle 
$\alpha=45^{\circ}$ and the rotation period  $P=0.1$~s.}
\label{polar}
\end{center}
\end{figure}
\begin{figure}

\begin{center}
\includegraphics{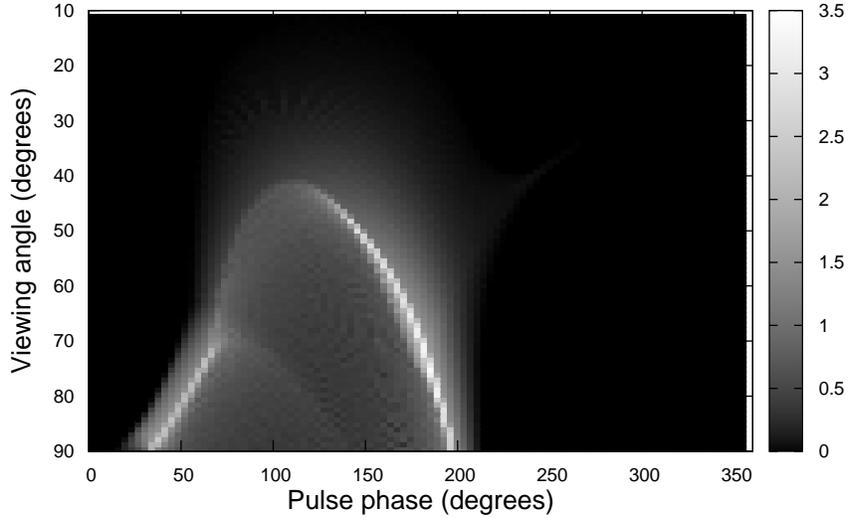}
\caption{Phase plot of the photons having energies larger than 100~MeV.
 The results are for $\alpha=60^{\circ}$, 
$B=3\times 10^{12}$~Gauss and $f_{zc}=0.1$. }
\label{map}
\end{center}
\end{figure}

\section{Results}
\label{result}
In this paper, we present the results of the two-layer outer gap model 
by using  the ratio of $h_1/h_2=0.95$  and 
the dimensionless charge density in the main region of $1-g_1=0.3$. 
The value  $h_1/h_2=0.95$  is chosen because 
 Wang et al. (2010) fitted the phase-averaged 
spectra of mature pulsars by using $h_1/h_2\sim 0.95$. The value  $1-g_1=0.3$
 is slightly larger than $1-g_1\sim 0.1$ used in  Wang et al. (2010), but 
reproduces a more consistent simulated population  with the $Fermi$ 
observations. The gap height $h_2$ and the charge density in screening region 
$g_2$ are calculated  from  equations~(\ref{fraction}) and~(\ref{condition}), 
respectively. 
 \subsection{Dependence on the inclination and viewing angles}
\label{dependency}
The dependence of the characteristics of the calculated 
$\gamma$-ray spectra on the inclination angle and the viewing 
angle are summarized in Figures~\ref{map}$\sim$\ref{indexMS}. 
Figure~\ref{map} shows the phase plot of the photons having the energies 
larger than 100~MeV. The result is for $\alpha=60^{\circ}$, 
$B_{12}=3\times 10^{12}$~G and $f_{zc}=0.1$.  

\subsubsection{$\gamma$-ray flux}
\label{flux}

The left panels in Figures~\ref{fluxCN} and~\ref{fluxMS} show the  
$\gamma$-ray flux ($\ge 100$~MeV) as a function of 
 the viewing angle $\zeta$ and of 
the inclination  angle $\alpha$.  The vertical line (y-axis) represents  
the fractional $\gamma$-ray flux, 
which is defined by the flux measured by   $f^3L_{sd}/d^2$ 
(c.f. Takata et al. 2011). Here we calculated 
 the rotation period from $B_s=3\times 10^{12}~$G and 
the gap fraction $f=f_{zc}$.  We can see  that 
the calculated flux tends to decrease as the line  of sight  approaches  to 
the rotation axis, where $\zeta=0^{\circ}$.  
In the sky map of  Figure~\ref{map}, we see that the large viewing 
angle ($\zeta \rightarrow 90^{\circ})$ can encounter more 
intense emission region, whereas the small viewing angle ($\zeta \ll 
90^{\circ}$) will encounter the less intense region 
 or even miss  the emission region. 
 With the outer gap geometry running  from 
the null charge surface to the light cylinder, 
the observer with a smaller viewing angle ($\zeta\ll 90^{\circ}$) 
may miss the emissions from higher latitudes ($z\ge h_2/2$)  
and measure the only 
emission from  the lower part of  the gap ($z\sim 0$). 
Because the accelerating electric field 
 vanishes at the lower boundary of the gap ($z=0$), the Lorentz factor 
of the accelerated particles and the resultant emissivity of the 
curvature radiation   around lower boundary of  the gap are significantly 
decreased.   Consequently,  a smaller viewing angle tends to measure 
 a smaller flux of the curvature radiation.  
In the left panels of Figures~\ref{fluxCN} and~\ref{fluxMS},
 we also see the tendency that the decrease 
of the fractional flux with the  decrease of the viewing angle is more gradual 
for larger inclination angle. This is because the emission from the outer gap 
with larger inclination angle covers  more wide region of the sky. 
These dependences  of the $\gamma$-ray flux on the viewing geometry    
predict that  the $Fermi$ has preferentially detected the pulsars with 
larger inclination angles  and  larger viewing angles near $90^{\circ}$ 
(c.f. section~\ref{popu}).

The right panels in Figure~\ref{fluxCN} and~\ref{fluxMS} summarize
 the dependence of the fractional $\gamma$-ray flux  on 
the gap fractional thickness. We can see that the  factional flux
 decreases as the gap fraction increases. 
This dependence  is related with the expansion 
of the outer gap in the azimuthal direction.
In the present calculation, we have  assumed  that the azimuthal 
expansion of the active gap is limited   as 
$f\le f_{max}(\phi)$ and $-90^{\circ}\le \phi\le 90^{\circ}$.  
As the dotted line in Figure~\ref{polar} shows, $f_{max}$ acquires  a 
maximum value near the magnetic meridian, where  $\phi=0^{\circ}$, and tends to 
decrease  as the azimuthal angle deviates from $\phi=0$.  This 
implies that the width of the active gap in the azimuthal direction narrows 
with the  increase of the gap fractional thickness, $f$. Consequently, 
the fractional  flux tends to decrease as the gap fraction  increases.

\subsubsection{Cut-off energy}
The cut-off energy is calculated as the peak energy in the spectral 
energy distribution of the emission from the main acceleration region. 
 Figures~\ref{ecutCN} and~\ref{ecutMS} show the 
dependence of the cut-off energies  for the CPs and for the MSPs, respectively, 
on the viewing geometry. The vertical axis represents the cut-off energy 
measured by  $3hc\Gamma_0^3/4\pi R_{lc}$, where 
$\Gamma_0=(3R_{lc}^2E_{||,0}/2e)^{1/4}$ with $E_{||,0}=f^2B(R_{lc})$. 
Figures~\ref{ecutCN} and~\ref{ecutMS} show that the cut-off energy 
decreases with the decrease of  the viewing angle. 
As we argued in section~\ref{flux}, 
a large viewing angle $\zeta\sim 90^{\circ}$ can encounter the emission 
from the strong accelerating electric field region at the 
middle  of the gap $(z\sim h_2/2$),  whereas 
the observer with a small viewing angle $(\zeta\ll 90^{\circ})$  measures
 the emission from the small electric field region near 
the lower boundary $(z\sim0$). 
As a result, the cut-off energy in the spectrum decreases with 
the decrease of the viewing angle. We can also find in Figures~\ref{ecutCN} 
and~\ref{ecutMS} that the fractional cut-off energy does not depend much 
on the fractional gap thickness, $f$.

\subsubsection{Photon index}
\label{index}
We applied the minimized-$\chi^2$ method  to fit the spectrum 
 between 100~MeV and the cut-off energy with a  single power low form. 
Figures~\ref{indexCN} 
and~\ref{indexMS} show the 
dependence of the photon index for the CPs and MSPs, respectively, on 
the viewing geometry. Our model predicts that the spectral shape 
is relatively soft with a photon index  
$p\sim 1.8-2$ for larger  viewing angle 
($\zeta \rightarrow 90^{\circ}$) and  hard with 
$p\sim 1.2-1.3$ for smaller viewing angle ($\zeta \ll 90^{\circ}$). 
In Figures~\ref{indexCN} and~\ref{indexMS}, 
the transition from soft to hard spectra 
occurs  in a   narrow range of the viewing angle. 
In the present two-layer model,  
the screening region and the main acceleration region produce 
the $\gamma$-ray photons with a typical energy of $\sim 100$~MeV and 
$\sim 1$~GeV, respectively. For the viewing angle closer to
 $\zeta\sim 90^{\circ}$,  because the observed $\gamma$-ray radiation 
consists of  the emissions from both main acceleration
 and screening regions, the emerging  spectrum becomes soft with 
 a photon index  of $p\sim 1.8-2$ above $100$~MeV.  
 For a smaller viewing angle, 
 on the other hand, because the emission 
of the screening region is missing, 
 the emission from  the only main acceleration region contributes to the spectrum.
 In such a case, the emerging spectrum  has 
a photon index  $p\sim 1.2-1.3$ above 100~MeV, which closes to 
a mono-energetic curvature spectrum. 
\begin{figure}
\begin{center}
\includegraphics{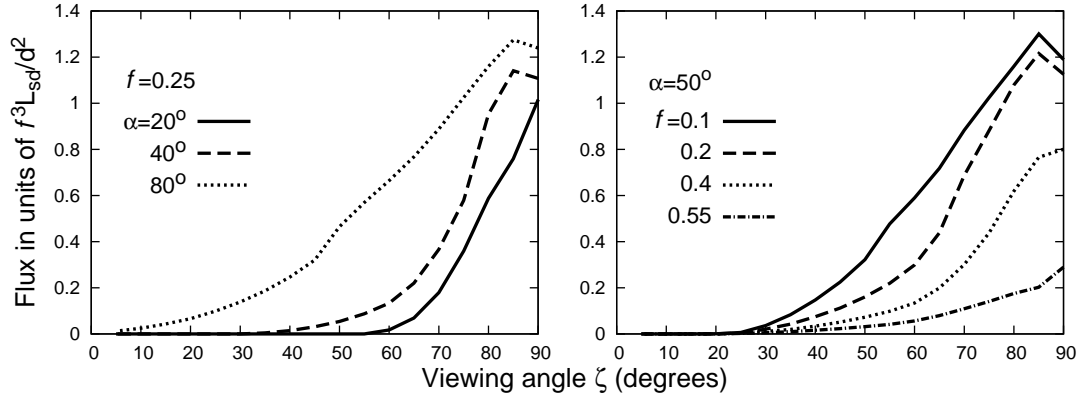}
\caption{The $\gamma$-ray flux ($>100$~MeV)  measured on the 
 as a function of the viewing angle.  The results are for the canonical
 pulsar with $B_s=3\times 10^{12}$~G and $f=f_{zc}$. 
Left:The $\gamma$-ray flux for  
the inclination angle is  $\alpha=20^{\circ}$ (solid line), 
$40^{\circ}$ (dashed line) and $80^{\circ}$ (dotted line). The results are 
for the gap fraction of $f=0.25$.  Right:The $\gamma$-ray flux for 
$f=0.1$~(solid line), 0.2~(dashed line), 0.4~(dotted line) and 
0.55~(dashed-dotted line). The results are for the inclination angle of 
$\alpha=50^{\circ}$. }
\label{fluxCN}
\end{center}
\end{figure}

\begin{figure}
\begin{center}
\includegraphics{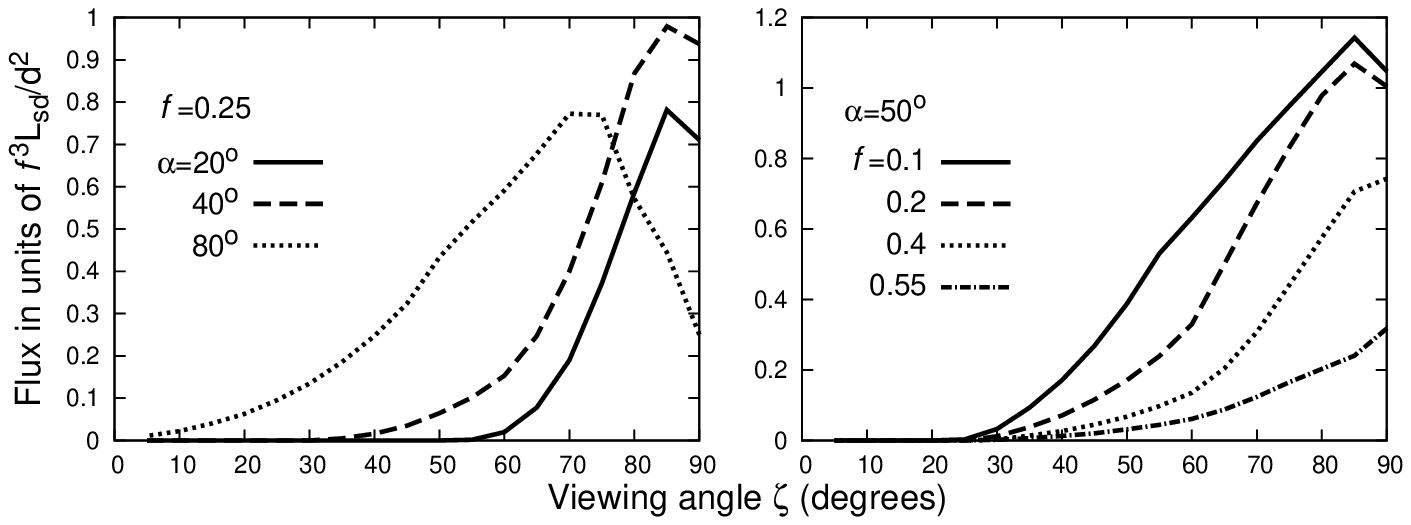}
\caption{The same with Figure~\ref{fluxCN}, but for the  millisecond pulsars
 with $B_s=3\times 10^{8}$~G. }
\label{fluxMS}
\end{center}
\end{figure}
\label{lastpage}

\begin{figure}
\begin{center}
\includegraphics{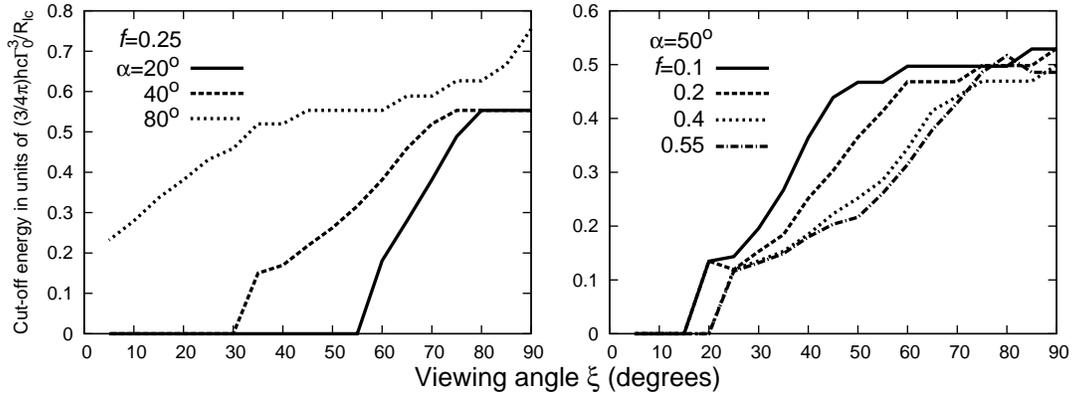}
\caption{Dependence of spectral cut-off energy  on the viewing geometry. 
The vertical line represent the cut-off energy in 
units of  $(3/4\pi)hc\Gamma_0^3/R_{lc}$, where
 $\Gamma_0=(3R_{lc}^2E_{||,0}/2e)^{1/4}$ with $E_{||,0}=f^2B(R_{lc})$.
 The results are for the canonical
 pulsar with $B_s=3\times 10^{12}$~G and $f=f_{zc}$. 
 The lines correspond to same cases as Figure~\ref{fluxCN}.}
\label{ecutCN}
\end{center}
\end{figure}

\begin{figure}
\begin{center}
\includegraphics{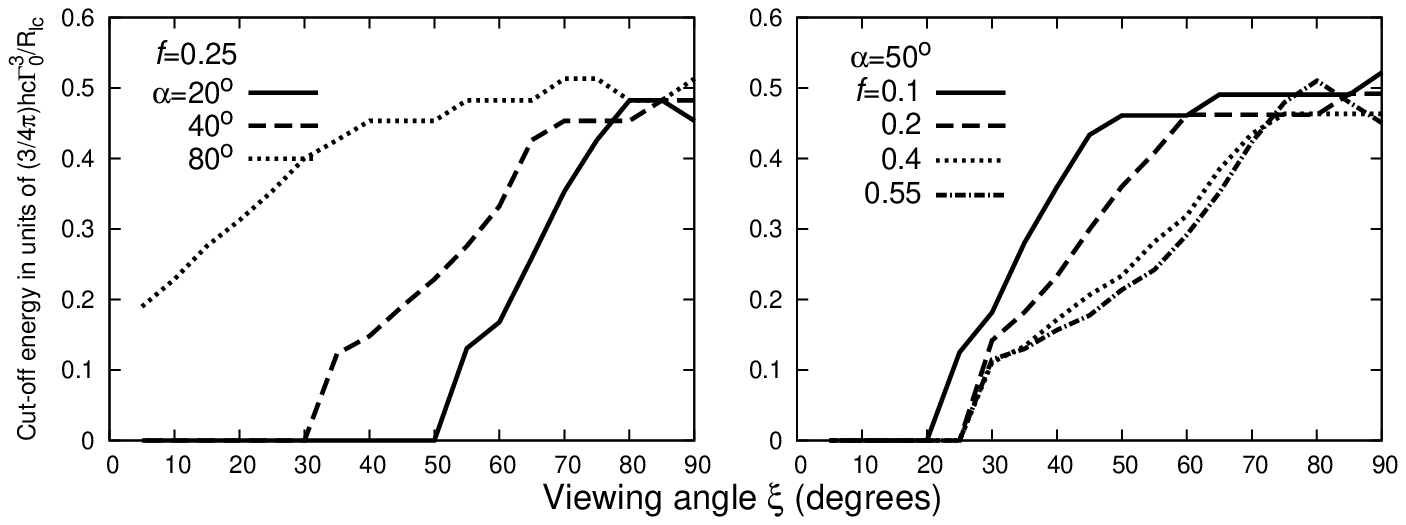}
\caption{The same with Figure~\ref{ecutCN}, but for the  millisecond pulsars
 with $B_s=3\times 10^{8}$~G. }
\label{ecutMS}
\end{center}
\end{figure}

\begin{figure}
\begin{center}
\includegraphics{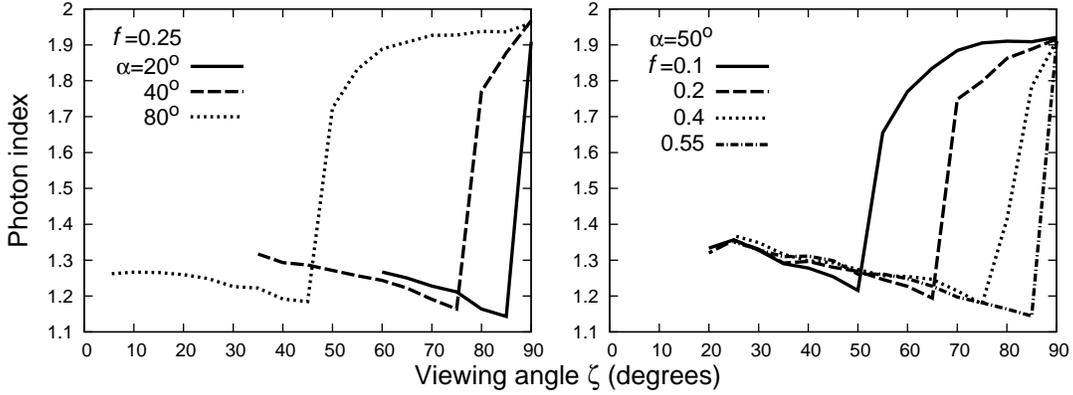}
\caption{Dependence of photon index   on the viewing geometry. 
The results are for the canonical 
 pulsar with $B_s=3\times 10^{12}$~G.  The lines correspond to same cases  
as Figure~\ref{fluxCN}.}
\label{indexCN}
\end{center}
\end{figure}

\begin{figure}
\begin{center}
\includegraphics{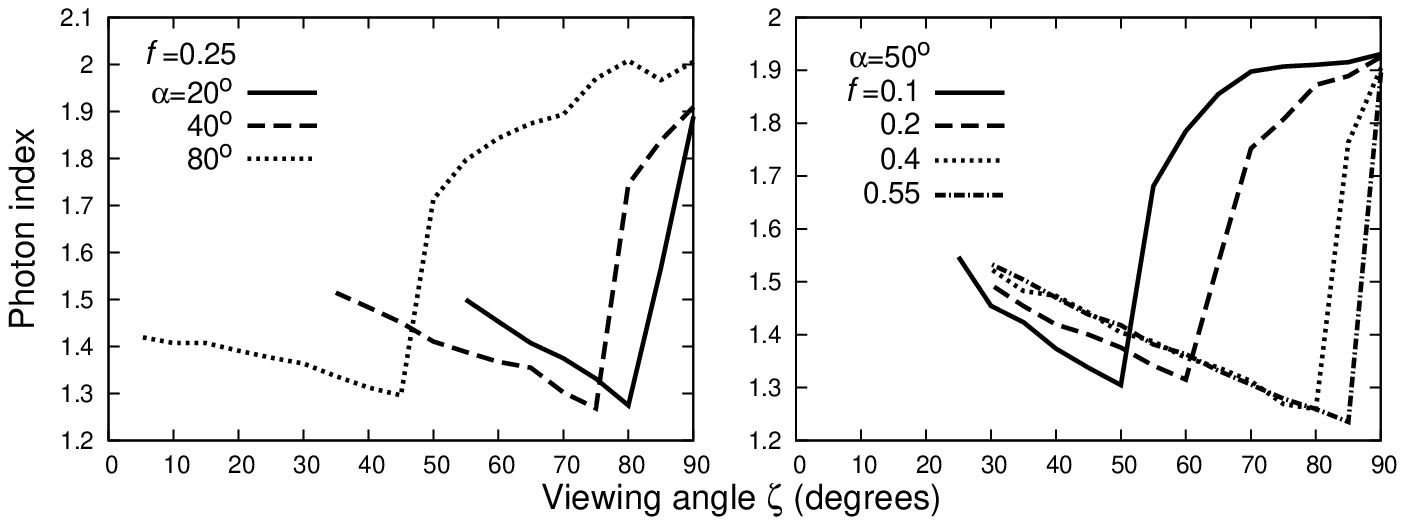}
\caption{The same with Figure~\ref{indexCN}, but for the  millisecond pulsars
 with $B_s=3\times 10^{8}$~G and $f=f_{zc}$. }
\label{indexMS}
\end{center}
\end{figure}

\subsection{Results of the Monte-Carlo simulation}
\label{monte}
\begin{table}
\begin{tabular}{ccccccc}
\hline
 & \multicolumn{2}{c}{six-months} & \multicolumn{2}{c}{five-years} & \multicolumn{2}{c}
{ten-years} \\CPs
 & $N_r$ & $N_g$ & $N_r$ & $N_g$ & $N_r$ & $N_g$ \\
\hline\hline
Ra. Sen. (x 1) & 40 & 39 & 56 & 138 & 59 & 182 \\
Ra. Sen. (x 2) & 51 & 34 & 77 & 123 & 81 & 163 \\
Ra. Sen. (x 10) & 72 & 28 & 130 & 90 & 145 & 120 \\
Beaming  & 76 & 28 & 155 & 79 & 177 & 102 \\
\hline
\end{tabular}
\caption{Population of simulated radio-selected ($N_r$) 
and $\gamma$-ray-selected ($N_g$) CPs for six-month, five-year and ten-year $Fermi$ observations. 
The first line; the results  for ten radio surveys 
 listed in table~1 of Takata et al. (2010a). The second and third lines are 
 the results 
with  the sensitivities increased  by a factor of two and ten, respectively, and  the bottom is the populations associated with only beaming effects 
of the radio emission.  }
\end{table}

\begin{table}
\begin{tabular}{ccccccc}
\hline
 & \multicolumn{2}{c}{six-months} & \multicolumn{2}{c}{five-years} & \multicolumn{2}{c}
{ten-years} \\ MSPs
 & $N_r$ & $N_g$ & $N_r$ & $N_g$ & $N_r$ & $N_g$ \\
\hline\hline
Ra. Sen. (x 1) & 10 & 52 & 14 & 200 & 16 & 284 \\
Ra. Sen. (x 2) & 16 & 48 & 26 & 190 & 29 & 274 \\
Ra. Sen. (x 10) & 45 & 32 & 82 & 152  & 94 & 227 \\
Beaming  & 106 & 11 & 321 & 41 & 438 & 62 \\
\hline
\end{tabular}
\caption{The same with Figure~1, but for the MSPs}
\end{table}

In this section, we present the results of the Monte-Carlo simulation, in 
which we assume the birth rates of the CPs and of the  MSPs are 
 $\sim$0.015 per year   and $\sim 9\times 10^{-6}$ per year, respectively. 
 We also assume that  the viewing angle and the inclination angle are 
randomly distributed. For the inclination angle close to $\alpha\sim
 90^{\circ}$, in which 
 the null charge points are  located  on or very close to the stellar 
surface,  it   may cause a numerical error when 
we search  the last-open field lines of the rotating vacuum field. 
To avoid it, we limit the inclination 
angle below $\alpha\le 85^{\circ}$.   Because a comparison 
between the  simulated and the observed 
distributions for the various properties (e.g. rotation 
period and magnetic field) of the $\gamma$-ray pulsars
 was done in Takata et al. (2011a,b), 
we focus on  the characteristics 
  of the  $\gamma$-ray radiations with the viewing geometry. 

For the sensitivity of the observations, we refer the $Fermi$ 
first pulsar catalog (Abdo et al. 2010a). For the 
radio-selected pulsars, the $Fermi$  archived  the sensitivity 
$F \sim 10^{-11}~\mathrm{erg/cm^2 s}$
 (or $\sim 3\times 10^{-11}~\mathrm{erg/cm^2 s}$) for the galactic latitudes 
$|b|\ge 5^{\circ}$ (or $<5^{\circ}$) with  the six-month long observations. 
For the CPs, the $Fermi$ six-month data allow us to detect  
 the pulsed period  by blind search, 
if the $\gamma$-ray flux  is larger  than 
$F\ge 2\times 10^{-11}$ for $|b|\ge 5^{\circ}$ and $F\ge 6\times 10^{-11}$
 for $|b|\le 5^{\circ}$. For the MSPs, because there are no 
such detections of the $\gamma$-ray-selected pulsar so far,
 we cannot simulate the $Fermi$
 sensitivity of the blind search. In this paper, therefore, we simulate 
the population of the $\gamma$-ray-selected MSPs with   
 the $Fermi$ sensitivity of the blind search of CPs.

\subsubsection{Viewing geometry}
Figures~\ref{axiCN} and~\ref{axiMS} plot the inclination angle ($\alpha$)
 and the viewing angle ($\zeta$) for the simulated 100 
canonical and millisecond 
$\gamma$-ray pulsars, respectively. The filled-circles and the boxes represent 
the radio-selected and $\gamma$-ray-selected $\gamma$-ray
 pulsars, respectively. For the CPs,  
we can see in Figure~\ref{axiCN} that the 
 radio-selected and $\gamma$-ray-selected pulsars distribute at 
 different region in $\alpha-\zeta$ plane. 
Specifically,  the  radio-selected CPs   group together 
around the line $\alpha=\zeta$. This is because we have assumed 
that the magnetic axis is the centre of the radio cone, 
indicating that the radio emission can be
 detected by observer with a viewing angle $\zeta\sim \alpha$. 
The scattering from the line $\alpha=\zeta$ is related with 
 the width of  the radio cone. For example, 
if we assume  a narrower cone  than that 
of equation~(\ref{kg}), the amplitude of the scattering is reduced.  
  Figure~\ref{axiCN} also shows that the 
detected $\gamma$-ray pulsars are mainly  distributed with 
a viewing angle $\zeta\sim 70-90^{\circ}$. These results are consistent with 
the results of Watters \& Romani (2011).

Unlike with the CPs, the distributions  of the radio-selected 
and $\gamma$-ray-selected  MSPs on the $\alpha$-$\zeta$ 
plane are overlapped  each other. This is because the width of the radio beam 
 described by equation~(\ref{kx})  covers almost whole sky. We note 
that most  $\gamma$-ray-selected MSPs  in the simulation 
 irradiate the Earth with the radio emissions, but the radio fluxes are  
lower than the sensitivity of the simulated radio
 surveys (c.f. Tables~1 and~2). 

In Figures~\ref{axiCN} and~\ref{axiMS}, we can see that 
no $\gamma$-ray pulsars  are detected with a  smaller inclination angles ($\alpha\ll 90^{\circ}$)  
and a smaller viewing angles ($\zeta\ll 90^{\circ}$). This is because the $\gamma$-ray flux decreases 
as the viewing angle and/or  the inclination angle decrease,  as we discussed 
 in section~\ref{dependency} 
(c.f. Figures~\ref{fluxCN} and~\ref{fluxMS}). Hence, 
 our simulation results predict that the pulsars with 
 larger inclination  and larger  viewing angles 
($\alpha,~\xi\rightarrow 90^{\circ}$) have 
 been preferentially  detected by the $Fermi$ six-month long observations.

\subsubsection{Population}
\label{popu}
Tables~1 and~2 show the simulated population of the CPs and MSPs, respectively, 
 with the $Fermi$ six-month, 
five-year and ten-year long  observations. Here we scale  
 the sensitivity of the $Fermi$ observations as  $\propto \sqrt{T}$, where 
$T$ is the length of the observation time. In addition, 
the second lines (``Ra. Sen. (x2)'') and the third lines (``Ra.Sen. (x10)'')
 in the tables show the results for the ten 
radio-surveys but we increase the sensitivities by a factor of two and of ten, 
respectively, and
 the fourth lines (``Beaming'') show the population associated with 
the only beaming effect of the radio emission.

 With the previous radio surveys (first line in Table~1), the present 
simulation shows that   
40 radio-selected  and 39 $\gamma$-ray-selected CPs  can be 
detected by  the $Fermi$ six-month observations, indicating that the present 
model predicts more $\gamma$-ray pulsars than the $Fermi$ observations 
($\sim 20$ for both radio-selected and
 $\gamma$-ray-selected $\gamma$-ray pulsars). 
 To explain the difference between the simulated and observed numbers,
 the several reasons will be expected; (1) the $Fermi$ 
 sensitivity will become much worse at  the Galactic plane, and 
(2) the $\gamma$-ray emissions from 
the pulsars will be missed by  source confusion with 
 the complex regions and  the  unresolved sources that are not modeled in the
 diffuse backgrounds.  The predicted number 10 of the radio-selected 
$\gamma$-ray MSPs is consistent with the observed number $9$. 
 The present model predicts 56 (or 59) radio-selected 
and 138 (or 182) $\gamma$-ray-selected CPs
 can be detected by the $Fermi$ with five-year (or ten-year) 
observations. For the MSPs, 14 (or 16) radio-selected $\gamma$-ray pulsars 
will be detected by the $Fermi$ with five-year (or ten-year) observations.

We see in the first lines of Tables~1 and~2  that  the simulated 
 numbers  of radio-selected $\gamma$-ray CPs and MSPs increase
only $\sim$20 and $\sim$10 sources, respectively, 
 over even  ten-year $Fermi$  observations.
 This implies that most presently
known radio pulsars ($\sim 2000$ for CPs and $\sim 80$ for MSPs) 
 might not be discovered by the  $Fermi$. 
For the $\gamma$-ray-selected pulsars, on the other hand,  
 the simulation predicts that the  $Fermi$ can 
detect about 140 CPs for about five-year observations. 
We note that the predicted number of the radio-loud (or radio-quiet) 
$\gamma$-ray pulsars really depends on the sensitivities  
of radio surveys, as Tables~1 shows; for example, 
the predicted number of the $\gamma$-ray-selected  CPs 
after ten-year of  the $Fermi$  observations decreases  from 182 to 120 if 
the sensitivities of the radio surveys increases by a factor of ten.  
  This indicates that a deep radio search may find  more  
radio-emissions from the $\gamma$-ray-selected  pulsars, 
such as LAT PSRs~J1741-2054 and~J2032+4127 (Camilo et al. 2009). 
Table~1 also shows if the sensitivities of the radio surveys increase  
by a factor of ten, the ratio of the radio-loud and radio-quiet  
$\gamma$-ray pulsars is almost determined by the beaming effect of 
the radio emissions. 

For the MSPs (Table~2), most  simulated pulsars are 
categorized as the $\gamma$-ray-selected pulsars 
with the previous sensitivities of the radio surveys, although 
the $Fermi$ has not confirmed the radio-quiet MSPs.  
We argue  that  it  may be difficult to identify  radio-quiet  
 MSPs, because   the detection of the rotation period 
by the $Fermi$  blind search is much harder than that of the CP. 
Furthermore, if the MSP is in binary system, the effects of the orbital
 motion on the observed rotation period make it even harder to confirm 
 the millisecond rotation period by the blind search.
 Takata et al. (2011b) discussed  that  
the  $\gamma$-ray-selected MSPs in the simulation correspond
 to  the $Fermi$ unidentified sources located at higher Galactic latitudes. 
 
We note that  the radio cones from MSPs are quite huge
 so that  most  MSPs irradiate the Earth with the radio emissions, as 
the bottom line in Table~2 shows. This implies that 
  more radio  MSPs  associated with the $Fermi$ 
unidentified sources will be detected by the future radio surveys, 
 such as  the discovery  of new 
20~radio MSPs associated with the $Fermi$ unidentified sources (Ray 2010;
Caraveo 2010; Ransom et al. 2011; Keith et al. 2011). 
As third line in Table~2 shows, however, even we drastically increase the radio 
sensitivity by a factor of 10, the number of radio-selected MSPs 
 detected by 10~year $Fermi$ observations
 can increase from  16 to 94, it is still 
much less than the expected 227 $\gamma$-ray-selected millisecond pulsars.
 Unless the $Fermi$  sensitivity of the blind search is improved, 
the most $\gamma$-ray MSPs will not be identified and will
 contribute to the $Fermi$ unidentified sources and/or the 
$\gamma$-ray background radiations. 

\subsubsection{$\gamma$-ray radiation characteristics}
Figures~\ref{fluxdis}-\ref{pindis} show the
 radiation characteristics ($\gamma$-ray luminosity, cut-off energy and photon index, respectively) versus the characteristics (the spin down power or 
the magnetic field strength at the light cylinder) for the $\gamma$-ray
 pulsars with the six-month observations. In those figures, 
we plot the $Fermi$ data with errors taken from Abdo et al. (2010a) 
and Saz~~Parkinson et al. (2010), and present the simulated pulsars
 in the sub-figures.  For the simulated pulsars, we randomly 
choose  200  simulated 
$\gamma$-ray pulsars (except for the $\gamma$-ray-selected MSPs).

In Figure~\ref{fluxdis},   the $\gamma$-ray luminosity is calculated from  
$L_{\gamma}=4\pi d^2 F_{\gamma,100}$, where $d$ is the distance, 
and  the histograms of the values of the $\gamma$-ray 
luminosity are  projected along the right-hand axis.  The solid  and
  dashed histograms represent the distributions
 for the simulated and observed $\gamma$-ray pulsars, respectively.
 The present model predicts 
that most  $\gamma$-ray CPs have a spin down power of 
 $L_{sd}\sim 10^{35-38}~\mathrm{erg/s}$ and a $\gamma$-ray luminosity  
of  $L_{\gamma}\sim 10^{34-36}$, while  MSPs have  a 
 $L_{sd}\sim 10^{33-35}~\mathrm{erg/s}$ and
 $L_{\gamma}\sim 10^{32.5-34.5}~\mathrm{erg/s}$, which are consistent with 
the $Fermi$ observations.

In Figure~\ref{fluxdis},  the spin down power  $L_{sd}$ and 
the $\gamma$-ray luminosity $L_{\gamma}$ of the simulated pulsars can be related as $L_{\gamma}\propto 
L_{sd}^{\beta}$ with $\beta\sim 0$ for $L_{sd}\ge 10^{35-36}~\mathrm{erg/s}$ 
and $\beta\sim 0.5$ for $L_{sd}\le 10^{35-36}~\mathrm{erg/s}$.   In 
the present emission model, the $\gamma$-ray luminosity is 
proportional to $L_{\gamma}\propto f^3L_{sd}$ as 
Figure~\ref{fluxCN} and~\ref{fluxMS} indicate (c.f. Takata et al. 2011b).
 The change of the slope  is caused by switching gap closure process
  between the photon-photon pair-creation process and the 
magnetic pair-creation process.
As the equations (\ref{fzccp})-(\ref{fm}) show, the gap fraction depends
 on  the rotation
 period and the magnetic field as  $f_{zc}\propto P^{26/21}B^{-4/7}$ 
for the photon-photon pair-creation  process 
and $f_{m}\propto P^{1/2}$ for the magnetic 
pair-creation process. These relations  imply  that 
the $\gamma$-ray luminosity depends on the spin down power as 
$L_{\gamma}\propto L_{sd}^{1/14}$ for the photon-photon pair-creation process
 and $L_{\gamma}\propto L_{sd}^{5/8}$ for the magnetic pair-creation process. 
Equating   $f_{zc}$~(\ref{fzccp}) and $f_{m}$~(\ref{fm}) corresponds to  
$L_{sd}\sim 10^{35-36}~\mathrm{erg/s}$. The change 
of the slope $\beta$ has been found by Wang et al (2010), 
who used the two-layer outer gap model to fit the  phase-averaged
 spectrum of the mature $\gamma$-ray pulsars observed by the $Fermi$. 

Although the general trend of the relation between the $\gamma$-ray 
luminosity and the spin down power 
 is explained with a simple form $L_{\gamma}\propto L_{sd}^{\beta}$, we can 
see in Figure~\ref{fluxdis} that 
 some simulation samples  deviate from the relation. For example, 
 some simulated pulsars with a spin down power of 
 $L_{sd}\sim 10^{36}~\mathrm{erg/s}$ have  a $\gamma$-ray luminosity of 
 $L_{\gamma}\sim 10^{32-33}~\mathrm{erg/s}$, which 
is about two or three  order smaller than the typical 
value of $L_{\gamma}\sim 10^{35}~\mathrm{erg/s}$.
 We emphasize that this low efficiency  of the $\gamma$-ray emission 
 is   mainly caused by the effects of viewing angle. The 
 pulsars lying on the relation $L_{\gamma}\propto L_{sd}^{\beta}$  
are observed with viewing angles of $\zeta\sim 90^{\circ}$, whereas 
those (``apparently'') low-efficient $\gamma$-ray pulsars are
 observed with smaller viewing angles.
  Although the flux depends on also the inclination angle, 
the flux  is more sensitive to the viewing angle  than the inclination angle, 
as Figures~\ref{fluxCN} and~\ref{fluxMS} show.
We note that  lower efficient $\gamma$-ray pulsars tend to  locate 
closer to the  Earth.  

Figure~\ref{ecutdis} shows  the cut-off energy versus 
 the magnetic field at the light cylinder, $B_{lc}$. 
 The symbols and the histograms correspond to same cases 
as Figure~\ref{fluxdis}.   As sub-figure in Figure~\ref{ecutdis} indicates, 
the present simulation  predicts  that most 
$\gamma$-ray CPs and MSPs have 
a magnetic field at the light cylinder of $B_{lc}\ge 10^{3}$~G 
and $B_{lc}\ge 10^4$~G, respectively,  and have  a cut-off energy 
smaller than $\sim2$~GeV. These features will be  consistent 
with the $Fermi$ observations. In the figure,  some simulated 
CPs  have  a cut-off 
energy significantly smaller than the typical value $E_{c}\sim 2$~GeV. 
As well as the case of the $\gamma$-ray luminosity, 
 this deviation from the typical value  is cause by the effects of the 
 viewing angle, as Figures~\ref{ecutCN} 
and~\ref{ecutMS} imply.  In Figure~\ref{ecutdis}, our simulation predicts 
 that  the typical cut-off energy (1-1.5~GeV) of the MSPs 
is smaller than that ($\sim 2$~GeV) 
of the CPs. It is difficult to  discuss  the difference in  the 
observed cut-off energies between the CPs and the MSPs,  because of 
the large observational errors.

Figure~\ref{pindis} represents the photon index versus  the 
spin down power. We can find that  the model 
distribution of the photon index has 
 two peaks at $p\sim 1.2-1.3$ and $p\sim 1.8-2$, and 
 the observed distribution (dashed
 histogram)  may also have   two peaks at $p\sim 1.3$ and $p\sim 1.7$. 
 With the present model, the hard component of   $p\sim 1.2-1.3$ corresponds 
 to the spectrum  associated with the emission from the only main  acceleration 
region, while   the soft component  $p\sim 1.8-2$ corresponds 
to the spectrum composed of the emissions from 
 both main and screening regions, as we discussed in section~\ref{index}. 
The present model does not predict  very hard spectrum with a index 
$p\le 1$, which has been indicated for  some $Fermi$ pulsars. 

Figures~\ref{fave} and~\ref{pave} represent the averaged  
{\it apparent} fractional thickness, which is defined by 
$f_a\equiv (4\pi d^2F_{\gamma}/L_{sd})^{1/3}$, and the photon index, respectively, as a function of the spin down power, and Figure~\ref{ecave} shows 
the averaged cut-off energy as a function of the magnetic field at the 
light cylinder. In the figures, 
the solid and dashed lines represent the results for the 
radio-selected and $\gamma$-ray-selected CPs, respectively.
 In addition, the thick and thin lines correspond to the results 
of the simulation and the $Fermi$ observations, respectively.  
In Figure~\ref{fave},  our model predicts  a tendency that 
   the apparent fractional thickness, $f_a$, tends to decrease
 with the increase of the spin down power. This is because the {\it true} 
fractional thickness, $f_{zc}$ or $f_m$,  
tends to decrease with the increase of the spin down power, that is,   
 $f_{zc}\propto L_{sd}^{-13/21}B^{1/21}$ from equation~(\ref{fzccp}) and 
$f_{m}\propto L_{sd}^{-1/8}B^{1/4}$ from equation~(\ref{fm}). 
Because the photon index of the spectrum tends to increase 
with the decrease of the fractional gap thickness 
as Figure~\ref{indexCN} shows, the averaged photon index 
in Figure~\ref{pave} increases with the spin down power.  We find that these behaviors are qualitatively consistent with the $Fermi$ observations. 
In Figure~\ref{ecave},  our model predicts that the typical 
cut-off energy does not depend much on the magnetic field 
strength at the light cylinder, while the present $Fermi$ data 
may  have a tendency that the cut-off energy increases with the magnetic field 
strength at the light cylinder. However, 
because the observational errors  are 
so large, a more deep observation to reduce the errors 
may be required to discuss the tendency.

In Figures~\ref{depenCN} and~\ref{depenMS}, we summarize the distributions 
of the radiation characteristics for the simulated canonical and millisecond 
$\gamma$-ray pulsars, respectively, including  both radio-selected and 
$\gamma$-ray-selected pulsars.   The solid and dashed lines are results 
for the simulated six-month and ten-year $Fermi$ observations, respectively.
 Comparing the solid and dashed lines, 
we find that the distributions do not depend much on the time span of 
the $Fermi$ observations. However, we note that a longer observation 
enables to detect $\gamma$-ray pulsars with smaller fluxes, which include 
$\gamma$-ray pulsars with   the viewing angle close to the rotation axis 
 (c.f. Figures~\ref{fluxCN} and~\ref{fluxMS}). 
 For the observer with a viewing angle  close to the rotation axis,  the photon 
index tends to be $p\sim1.2-1.3$, 
 as Figures~\ref{indexCN} and~\ref{indexMS} show. 
Therefore, our model prediction is that  
a longer $Fermi$ observation will detect 
more $\gamma$-ray pulsars with 
photon indexes  $p\sim 0.12-0.14$,
 as right panels in Figures~\ref{depenCN} and~\ref{depenMS} 
indicate.

\subsubsection{Pulse profiles}
\label{pulse}
Figures~\ref{pulser}-\ref{pulsegMS}  present the calculated pulse
 profiles for the different type of the $\gamma$-ray pulsars. 
For each type of the $\gamma$-ray pulsars, 
64 samples  are randomly chosen to present the model prediction on
 the statistical distribution  of the morphology of the pulse profiles. 
From the left to right panels and from the top to bottom panels in the figures, 
the inclination angle  increases.  
In principle, we can quantitatively  compare the simulated 
morphology  of the pulse profiles in  Figures~\ref{pulser}-\ref{pulsegMS}
 (e.g. phase-separation, number of peak) 
 with the $Fermi$ observations. However, we would like to point out 
that it is very difficult to quantify  the morphology to compare with the 
observations.  
In  Figure~\ref{pulser}, for example, the pulse  profile  represented 
in upper-left conner is indeed  a double peak structure with the narrow 
phase-separation between two peaks. In the  observations, however, 
the identification of the double peak structure with such  narrow peak 
separation really depends on source counts and  timing ephemerides 
of the pulsars. In   Figure~\ref{pulseg}, 
the simulated pulse profile presented  
at the seventh-line and the first-column ($\alpha=74.2^{\circ}$ and 
$\zeta=67.5^{\circ}$) has the first peak much smaller 
than the second peak. In the observations, the detection  
of such a small peak will depend on the strength of the 
background radiation.  Moreover, 
because the intensity of pulse peak will depend on 
the  modeling of the gap structure, it is difficult to discuss 
the distribution of the 
intensity ratio of the first and the second peaks with the present simple 
three-dimensional model. In the present paper, therefore, we avoid 
a quantitative discussion for the morphology of the pulsed profile.
 Instead, we can only present a qualitative comparison.

 We note  that the peaks emerging in the calculated pulse profiles
are caused by the so-called caustic effect (Romani and Yadigaloglu 1995; 
Cheng, Ruderman and Zhang 2000; Dyks, Harding and Rudak 2004), 
in which more photons are observed at narrow width of the rotation phase due to
the special-relativistic effects, i.e.  the aberration of the
emission direction and photon's travel time. In such a case, 
the pulse phase and the number of the main peak are not affected much by 
  modeling of the gap structure, whereas 
 the intensity of the peak depends on the gap structure.  

We can see in  Figures~\ref{pulser} and~\ref{pulseg} that 
 the  pulse profiles of the simulated  CPs (in particular,  
$\gamma$-ray-selected CPs) are described by 
 double peak structure rather than single peak profile. 
 This result may explain the tendency  that  most   canonical 
$\gamma$-ray pulsars observed by the $Fermi$ show the double peak structure. 
In the present simulations, we have predicted that  the $\gamma$-ray 
pulsars measured from  $\zeta\sim 90^{\circ}$ are 
preferentially detected in the simulation (c.f. section~\ref{dependency}).
With the viewing angle closer to $\zeta\sim 90^{\circ}$, the 
calculated pulse profile tends to have the double peak structure. 

For the radio-selected canonical $\gamma$-ray pulsars, 
although the double peak structure are more or less common feature, 
 single pulse structure or the double peak 
structure with a narrow phase separation stands out for pulsars 
 with inclination angles  $\alpha\sim 40^{\circ}-50^{\circ}$, as we can see in
 Figure~\ref{pulser}.  This is because  the 
simulated radio-selected canonical pulsars 
have the viewing angle similar to the inclination 
angle (c.f. Figure~\ref{axiCN}), $\zeta\sim \alpha$. 
With the present outer gap geometry,  
the pulse profile of the 
smaller viewing angle ($\zeta\ll 90^{\circ}$) tends  to have 
 single peak or  double-peak with a narrow phase separation, as the 
phase plot  of Figure~\ref{map} shows. 
 
Comparing the pulse profiles of the CPs (Figures~\ref{pulser} and~\ref{pulseg})
 and of the MSPs (Figures~\ref{pulserMS} and~\ref{pulsegMS}), 
  one may see that  the pulse profile   
with single peak  is  more common for the  MSPs than  the  CPs. 
 In the present model, the  fractional gap thickness $f_{ZC,MSP}$ 
of equation~(\ref{fzcmp}) for the MSPs tends 
to be larger than $f_{ZC, CP}$ of  equation~(\ref{fzccp}) for the CPs. 
With a larger  fractional gap thickness,
  the emission from the higher altitude
(larger $z$ in units of the light radius)  contributes to 
 the  pulse profile.  Because the emission from the higher altitude 
produces two caustic peaks with a narrower phase separation,  the calculated 
 pulse profiles of the MSPs have  single peak or double 
peak structure with the narrow phase separation  more than that of the CPs.
 
\section{Summary and Discussion}
\label{summary}
In this paper, we have applied the so called two-layer outer gap model for 
the $\gamma$-ray radiations from the pulsars, and have focused  
on  the dependence of the $\gamma$-ray radiation 
characteristics on the inclination angle and the viewing angle. We showed 
that the $\gamma$-ray flux and the spectral cut-off energy decreases   as 
the viewing angle deviates from  $\zeta=90^{\circ}$. 
The spectrum above 100~MeV 
becomes soft with a photon index  $p\sim 1.8-2$ for the 
 observer  with a larger viewing  angle  ($\zeta \rightarrow 90^{\circ}$),  
whereas it becomes hard with a index  $p\sim 1.2-1.3$ 
for the observer with a smaller viewing angle ($\zeta\ll 90^{\circ}$).  
 The spectrum with a 
 photon index $p\sim 1.8-2$ consists of the emissions 
from both main acceleration and screening regions, 
while $p\sim 1.2-1.3$ corresponds to the emission from the only  
main acceleration region.

We  have developed   the Mote-Carlo simulation 
 for the population of $\gamma$-ray emitting pulsars. The our simulation 
predicts that 56  (or 59) radio-selected,  138 (or 182) 
$\gamma$-ray-selected canonical $\gamma$-ray pulsars and 14 (or 16)
 radio-selected $\gamma$-ray MSPs can be detected 
by five-year  (or ten-year) $Fermi$ observations. 
 Even we drastically increase the radio 
sensitivity by a factor of ten, the most simulated  MSPs are expected as 
 the $\gamma$-ray selected MSPs, which will contribute as the $Fermi$ 
unidentified sources and/or $\gamma$-ray background radiations. 
For the viewing geometry,   our simulation predicts  
that the radio-selected CPs have  the inclination angle ($\alpha$) 
and the viewing angle ($\zeta$) of   $\alpha\sim \zeta$, and  that most 
$\gamma$-ray  pulsars  detected by the $Fermi$ have been measured with 
 a  viewing angle of $\zeta=70^{\circ}\sim 90^{\circ}$.  We demonstrated 
 that the spin down power  $L_{sd}$ and 
the $\gamma$-ray luminosity $L_{\gamma}$ of the simulated pulsars is 
 related as $L_{\gamma}\propto 
L_{sd}^{\beta}$ with $\beta\sim 0$ for $L_{sd}\ge 10^{35-36}~\mathrm{erg/s}$
  and $\beta\sim 0.5$ for 
$L_{sd}\le 10^{35-36}~\mathrm{erg/s}$. The change of the 
slope $\beta$ is associated with the switching of the gap closure mechanism 
from the photon-photon pair-creation process 
of  $L_{sd}\ge 10^{35-36}~\mathrm{erg/s}$ to 
the magnetic pair-creation process of $L_{sd}\le 10^{35-36}~\mathrm{erg/s}$.
  We  showed that the distribution of the photon index 
of the simulated $\gamma$-ray pulsars  
has  two peak at $p\sim 1.8-2$ and $p\sim 1.2-1.3$, which may explain 
the observed distribution. We expect 
 that more  $\gamma$-ray pulsars with the hard spectrum  of $p\sim 1.2-1.3$
  will be detected by the future $Fermi$ observations. 
 For the pulse profiles,  the present simulation will explain 
the observational tendency  that
 most canonical $\gamma$-ray pulsars detected by the $Fermi$ 
 have the  pulse profiles with the double peaks. 
 The our model expects  that single pulse profile or 
the double pulse profile with a narrower phase separation is  more 
common for  MSPs than CPs.

The present model can be used to diagnose the viewing geometry 
of the $\gamma$-ray pulsars.  For example, the present model predicts that 
 most  present $\gamma$-ray pulsars detected by the $Fermi$ will have 
the viewing angle, $\zeta\sim 70^{\circ}-90^{\circ}$, 
and hence the pulse profile with the double peaks is 
more common than that with the single peak. On the other hand, we expect that  
PSR~J0659+1414, which is known as the extremely  low efficient
 $\gamma$-ray pulsar (Abdo et al. 2010c),  has a unique  viewing geometry.
  PSR~J0659+1414 is the radio-loud 
$\gamma$-ray pulsar and its  spin down power is 
$L_{sd}=3\times 10^{34}$~erg/s. Interestingly, 
the $\gamma$-ray emissions from PSR~J0659+1414 has been observed
 with a  $\gamma$-ray luminosity, 
$L_{\gamma}\sim 3\times 10^{32}~\mathrm{erg/s}$, which 
  is about two order smaller than the typical value for the pulsars 
with a similar spin down power. 
To explain the observed low efficiency of the $\gamma$-ray radiation, 
 our simulation  predicts  that  the inclination angle and 
the viewing angle are relatively small, say  
$\alpha\sim \zeta\sim 40^{\circ}-50^{\circ}$, and hence the apparent
 efficiency  is  significantly reduced  from 
the typical value (c.f. Figure~\ref{fluxCN}). 
 PSR~J0659+1414 is also known that the spectral cut-off energy, 
$E_{c}\sim 0.7$~GeV, is suggestively smaller than the typical value 
of  $E_c\sim 2$~GeV. This behavior is also able to be explained by the 
 effects of the viewing geometry, as Figure~\ref{ecutCN} indicates. 
 Furthermore, our simulation expects that the pulsars measured as  the low efficient $\gamma$-ray 
pulsars are  located closer to  Earth. 
  In fact, PSR~J0659+1414 with $d\sim 0.28$~kpc is 
one of the nearest $\gamma$-ray pulsars. 
Finally, the observed broad single pulse profile of PSR~J0659+1414 is also 
expected by  the present calculation.  As discussed in section~\ref{pulse} (Figure~\ref{pulser}), 
 the single pulse structure or  double peak 
structure with a narrow phase separation is more common for 
the radio-selected  canonical $\gamma$-ray pulsar with an  
inclination angle $\alpha\sim 40^{\circ}-50^{\circ}$. 
On these ground,   
 the unique radiation properties  of   PSR~J0659+1414 can be explained, 
if the $Fermi$  has  measured  the $\gamma$-ray emission from the  our gap
 with the viewing geometry $\alpha\sim \zeta\sim  40^{\circ}-50^{\circ}$. 
 It is expected that the population of  the 
low-efficient $\gamma$-ray pulsars will be increased by the future 
$Fermi$ observations.  The diagnose its $\gamma$-ray efficiency, 
 pulse profile and cut-off energy
 will provide  a strong constraint on the emission models.

 We thank   A.H. Kong, C.Y.~Hui, B.~Rudak,
 Lupin-C.C. Lin, M.Ruderman, R.E. Taam and  
S.Shibata for the useful discussions. 
We express our appreciation to an anonymous referee for  useful 
 comments. We also thank the  Theoretical Institute
for Advanced Research in Astrophysics (TIARA) operated under the Academia
Sinica Institute of Astronomy and Astrophysics, Taiwan,
which  enable author (J.T.) to use the PC cluster at TIARA.
KSC is supported by a 2011  GRF grant of the Hong Kong SAR
Government entitled ``Gamma-ray Pulsars''.

\begin{figure}
\begin{center}
\includegraphics{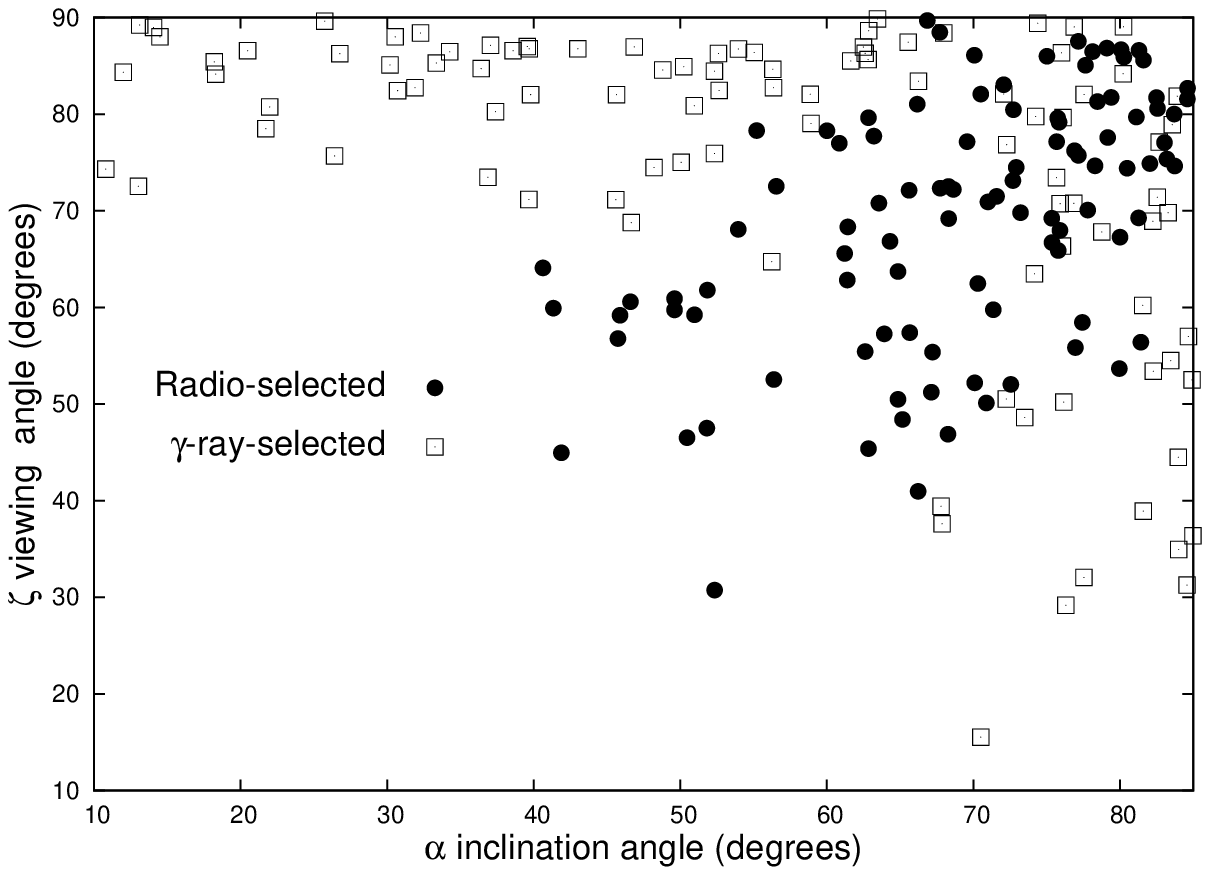}
\caption{The rotation axis $\alpha$ and the viewing angle $\zeta$ for 
the simulated canonical $\gamma$-ray pulsars. Filled-circles:Radio-selected 
pulsars. Boxes:$\gamma$-ray-selected pulsars.}
\label{axiCN}
\end{center}
\end{figure}

\begin{figure}
\begin{center}
\includegraphics{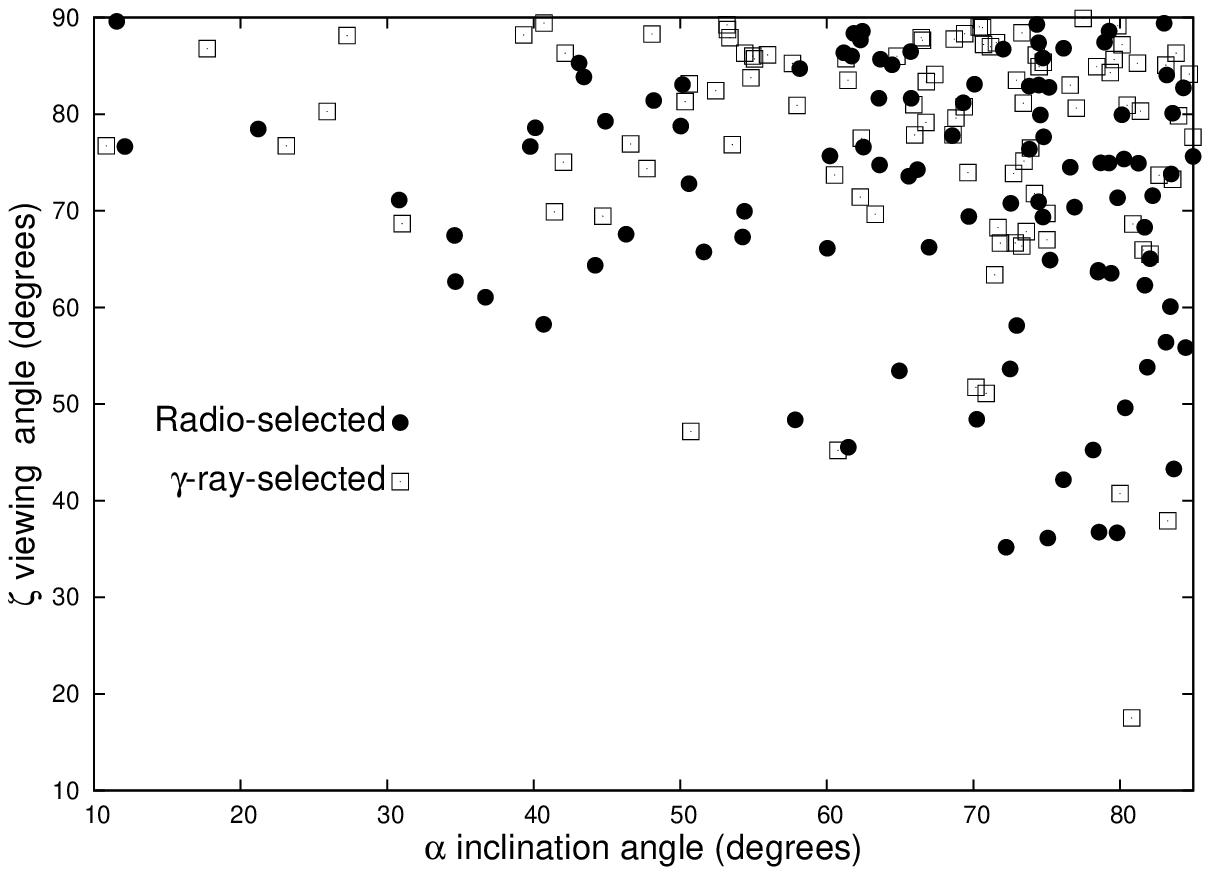}
\caption{The same with Figure~\ref{fluxCN}, but for the  millisecond pulsars.}
\label{axiMS}
\end{center}
\end{figure}

\begin{figure}
\begin{center}
\includegraphics{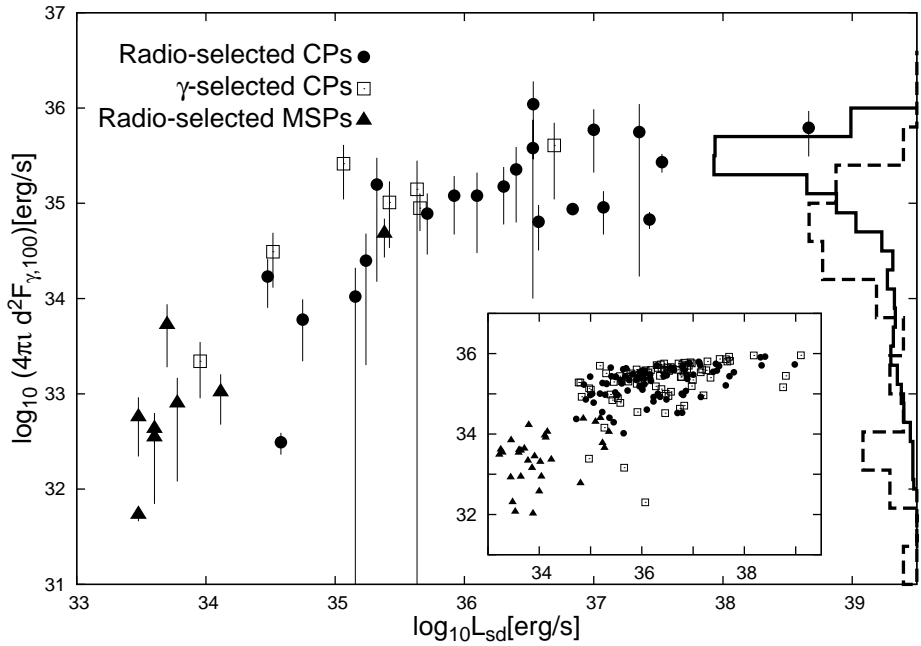}
\caption{ The 
$\gamma$-ray luminosity, $L_{\gamma}=4\pi d^2 F_{\gamma,100}$, 
 versus the spin down luminosity, $L_{sd}$, 
as result of  th six-month long observation. 
The $Fermi$ data with errors are taken from Abdo et al. (2010a) 
and Saz~~Parkinson et al. (2010), and  the 200 samples of the simulated 
pulsars are plotted  in the sub-figure. 
 in the sub-figures. Filled-circles:Radio-selected CPs. 
Square:$\gamma$-ray-selected CPs. Filled-triangles:Radio-selected MSPs.  
The histograms of the values of the $\gamma$-ray 
luminosity are  projected along the right-hand axis. 
 The solid and dashed histograms are results for  simulated and observed 
pulsars, respectively.}
\label{fluxdis}
\end{center}
\end{figure}

\begin{figure}
\begin{center}
\includegraphics{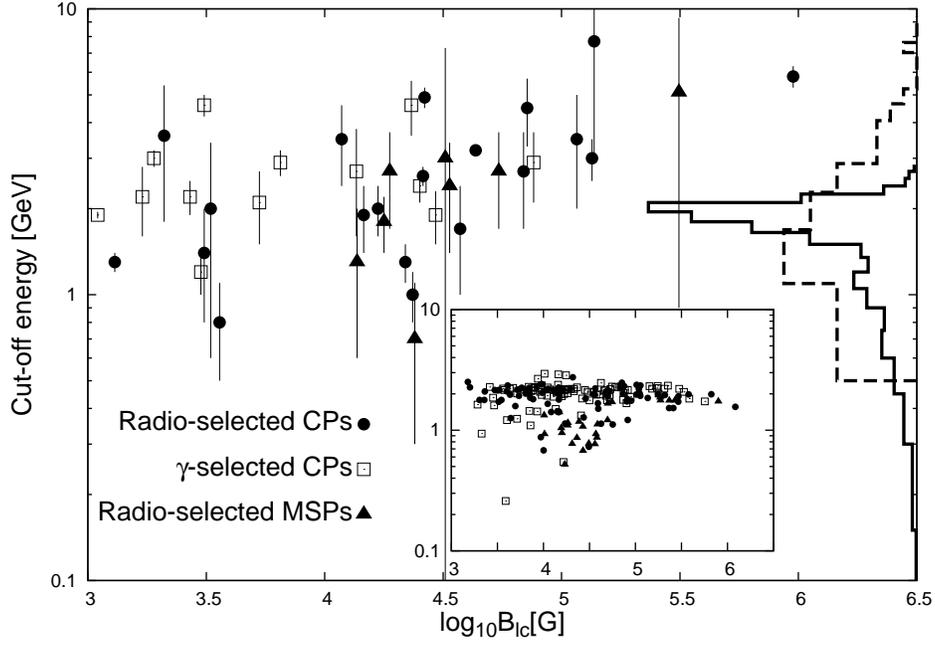}
\caption{The cut-off energy versus 
the magnetic field at the light cylinder, $B_{lc}$. 
 The symbols and histograms correspond to same cases as 
Figure~\ref{fluxdis}. }
\label{ecutdis}
\end{center}
\end{figure}

\begin{figure}
\begin{center}
\includegraphics{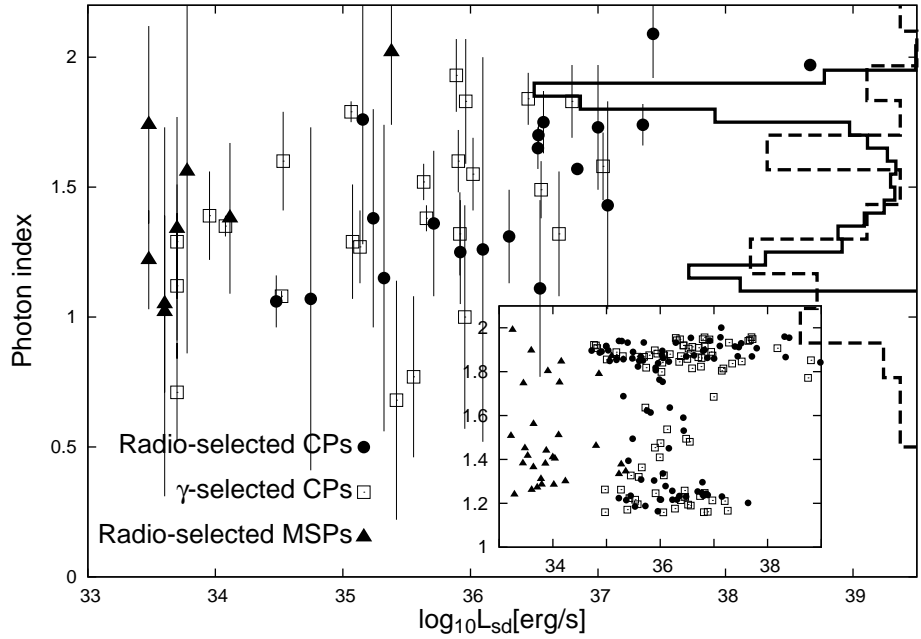}
\caption{The photon index versus  
the spin down luminosity.  The symbols and histograms correspond to same cases 
as Figure~\ref{fluxdis}. }
\label{pindis}
\end{center}
\end{figure}

\begin{figure}
\begin{center}
\includegraphics[height=6cm,width=6cm]{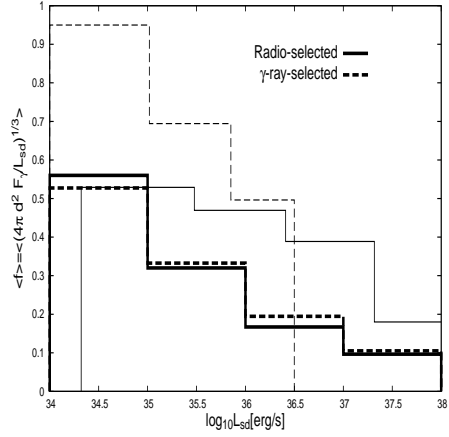}
\caption{Averaged value  of the {\it observed} 
fractional gap thickness, $f_{o}\equiv  (4\pi d^2 F_{\gamma}/L_{sd})^{1/3}$, 
of the detected  $\gamma$-ray pulsars as function of the spin down power.  
  The solid and dashed histograms  
show  the  results of the radio-selected and $\gamma$-ray 
selected pulsars, respectively.
 The thick and thin lines correspond to the simulation 
and $Fermi$ observation, respectively.}
\label{fave}
\end{center}
\end{figure}

\begin{figure}
\begin{center}
\includegraphics[height=6cm,width=6cm]{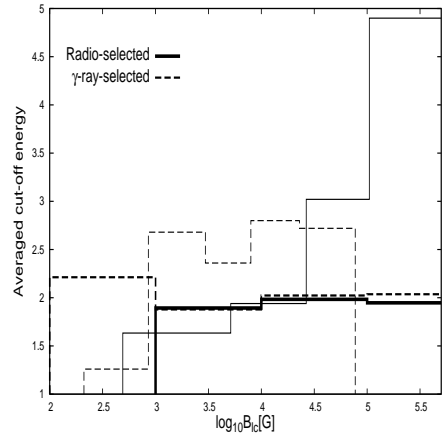}
\caption{Averaged value  of the cut-off energy of  
 the detected  $\gamma$-ray pulsars as function of the spin down power.  
The histograms  correspond to same cases 
as Figure~\ref{fave}.}
\label{ecave}
\end{center}
\end{figure}

\begin{figure}
\begin{center}
\includegraphics[height=6cm,width=6cm]{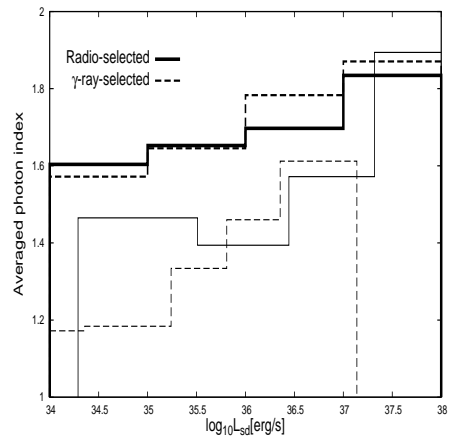}
\caption{Averaged value  of the photon index of  
 the detected  $\gamma$-ray pulsars as function of the spin down power.  
The histograms  correspond to same cases 
as Figure~\ref{fave}.}
\label{pave}
\end{center}
\end{figure}

\begin{figure}
\begin{center}
\includegraphics[height=6cm,width=15cm]{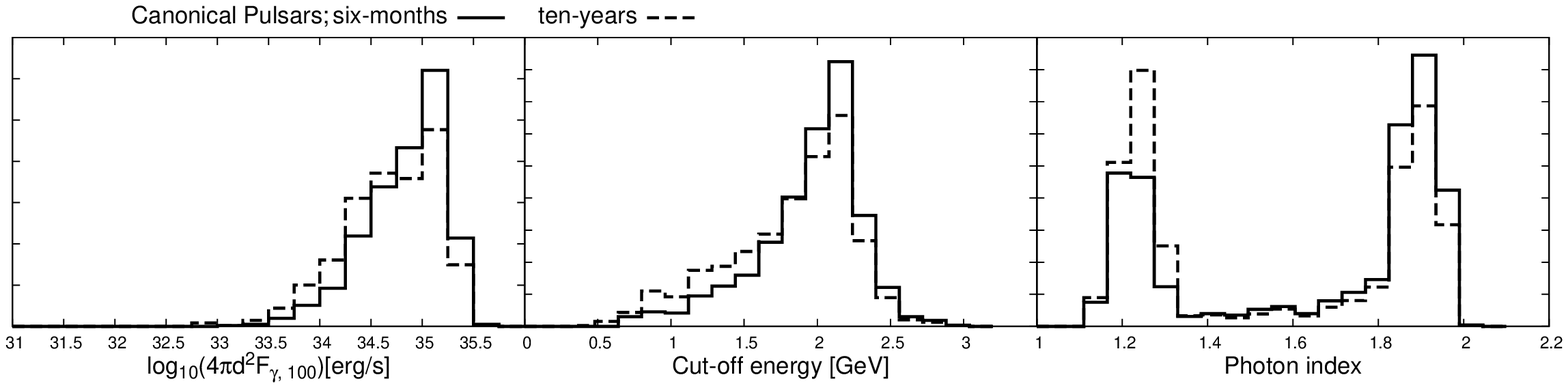}
\caption{The simulated distributions for the canonical 
pulsars. Left:Luminosity. Middle:Cut-off energy. Right:Photon index. The 
solid and dashed lines are results for simulated 
 six-month and ten-year $Fermi$ observations, respectively.}
\label{depenCN}
\end{center}
\end{figure}

\begin{figure}
\begin{center}
\includegraphics[height=6cm,width=15cm]{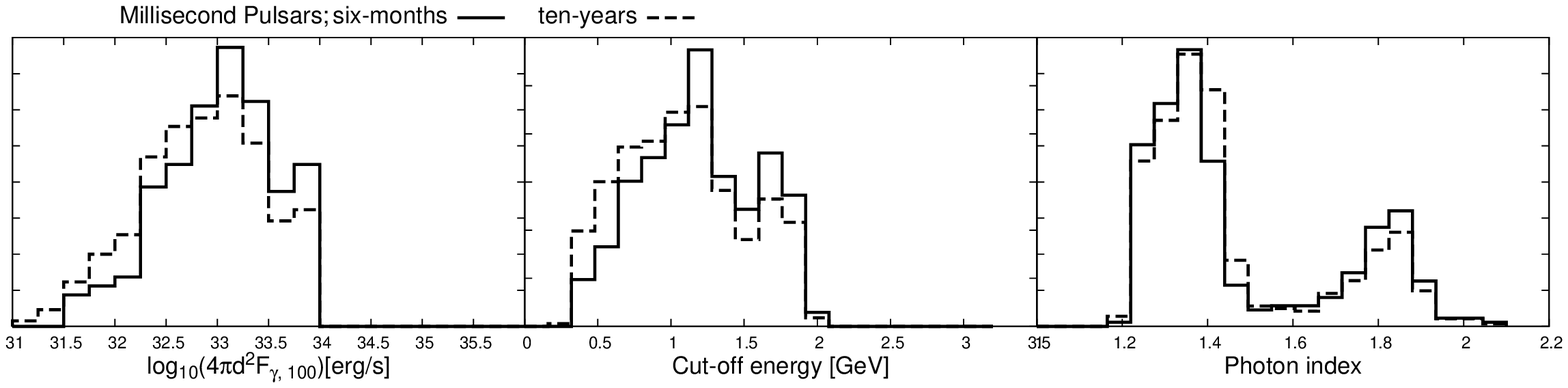}
\caption{The same with Figure~\ref{depenCN}, 
but for the  millisecond pulsars.}
\label{depenMS}
\end{center}
\end{figure}

\begin{figure}
\begin{center}
\includegraphics[height=12cm,width=17cm]{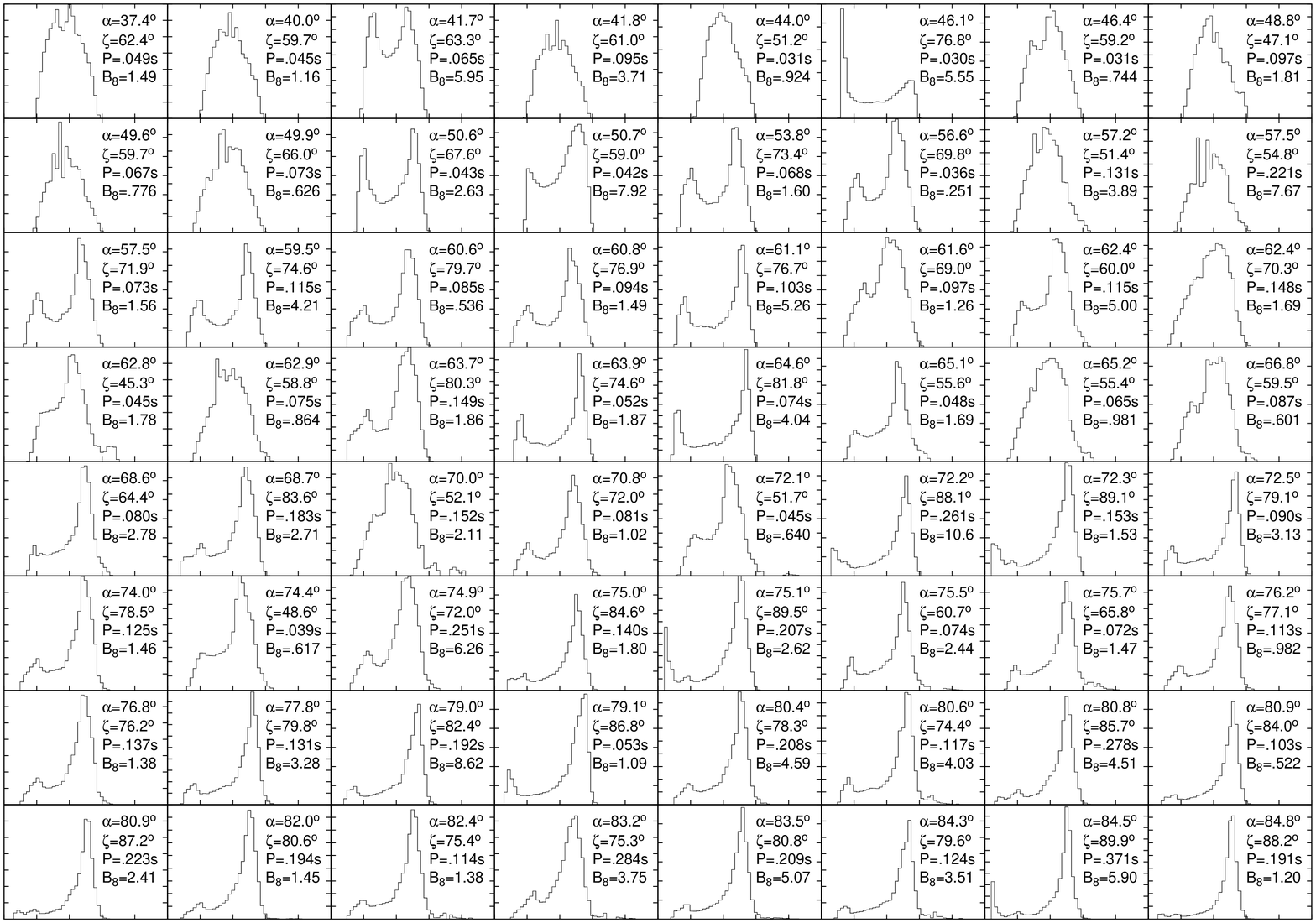}
\caption{The pulse profiles of 64 samples of the simulated 
radio-selected $\gamma$-ray pulsars.}
\label{pulser}
\end{center}
\end{figure}

\newpage

\begin{figure}
\begin{center}
\includegraphics[height=12cm,width=17cm]{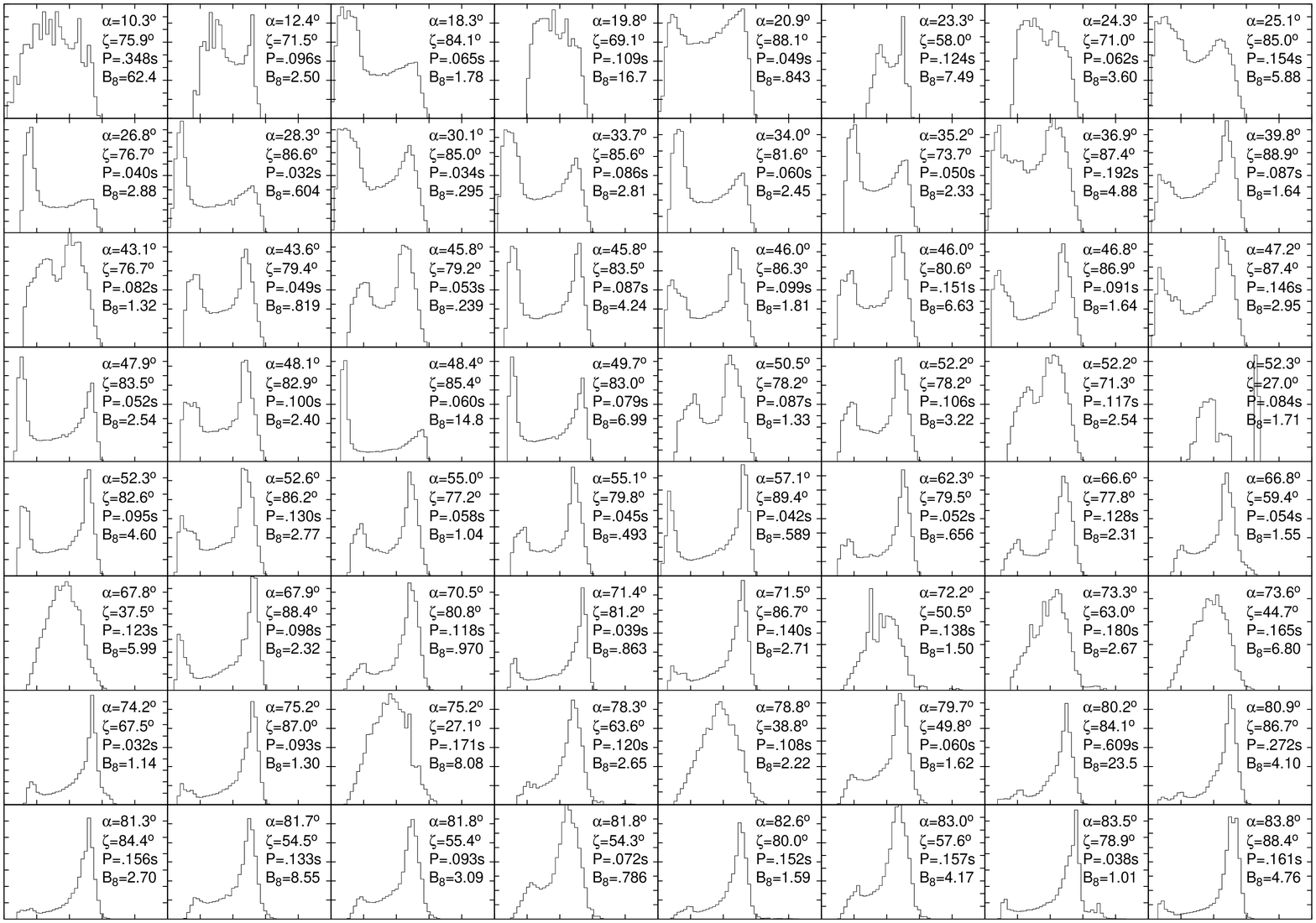}
\caption{The pulse profiles of 64 sample of the simulated $\gamma$-ray 
selected canonical pulsars.}
\label{pulseg}
\end{center}
\end{figure}

\begin{figure}
\begin{center}
\includegraphics[height=12cm,width=17cm]{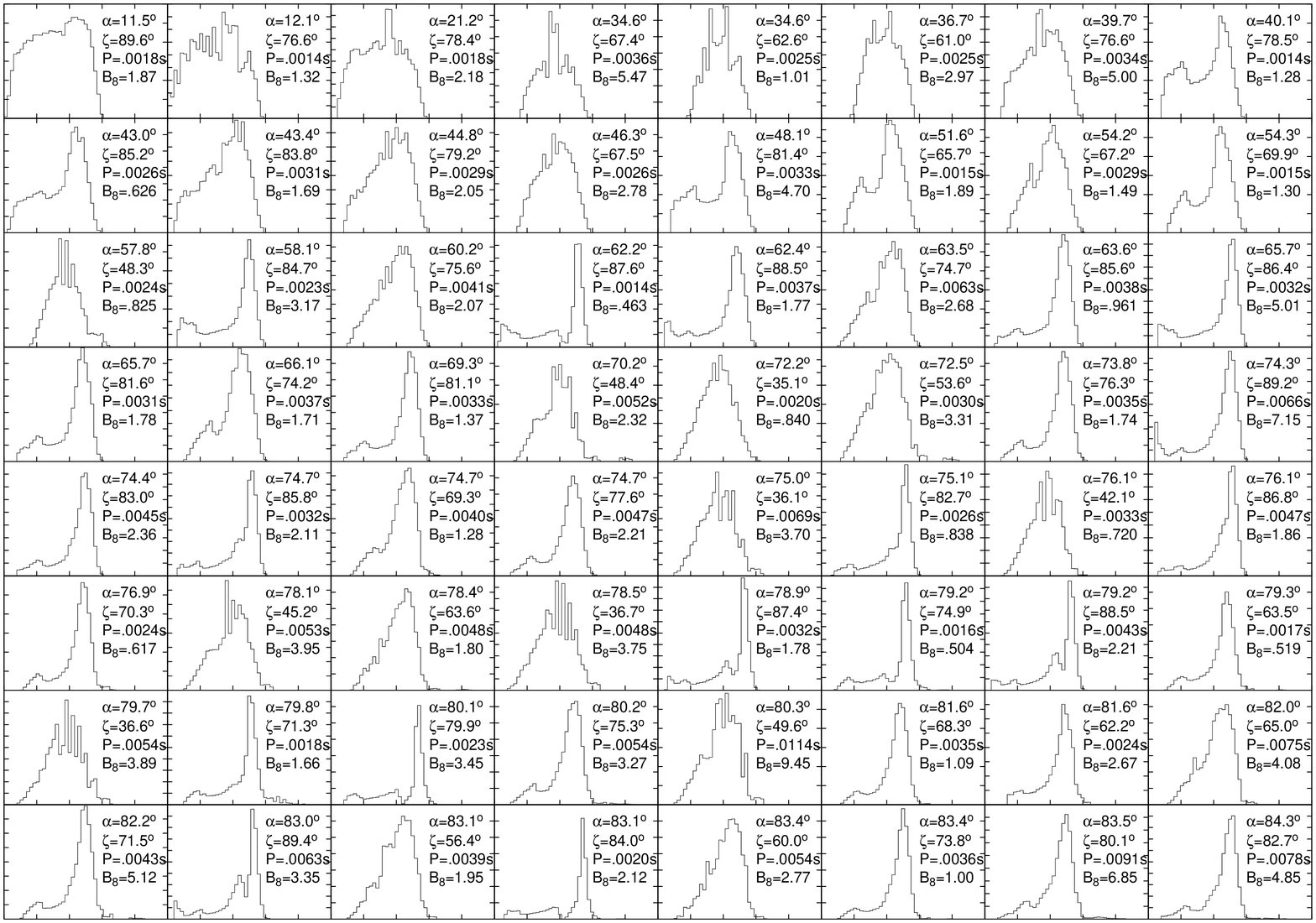}
\caption{The pulse profiles of 64 sample of the simulated radio-selected 
millisecond  $\gamma$-ray pulsars.}
\label{pulserMS}
\end{center}
\end{figure}

\begin{figure}
\begin{center}
\includegraphics[height=12cm,width=17cm]{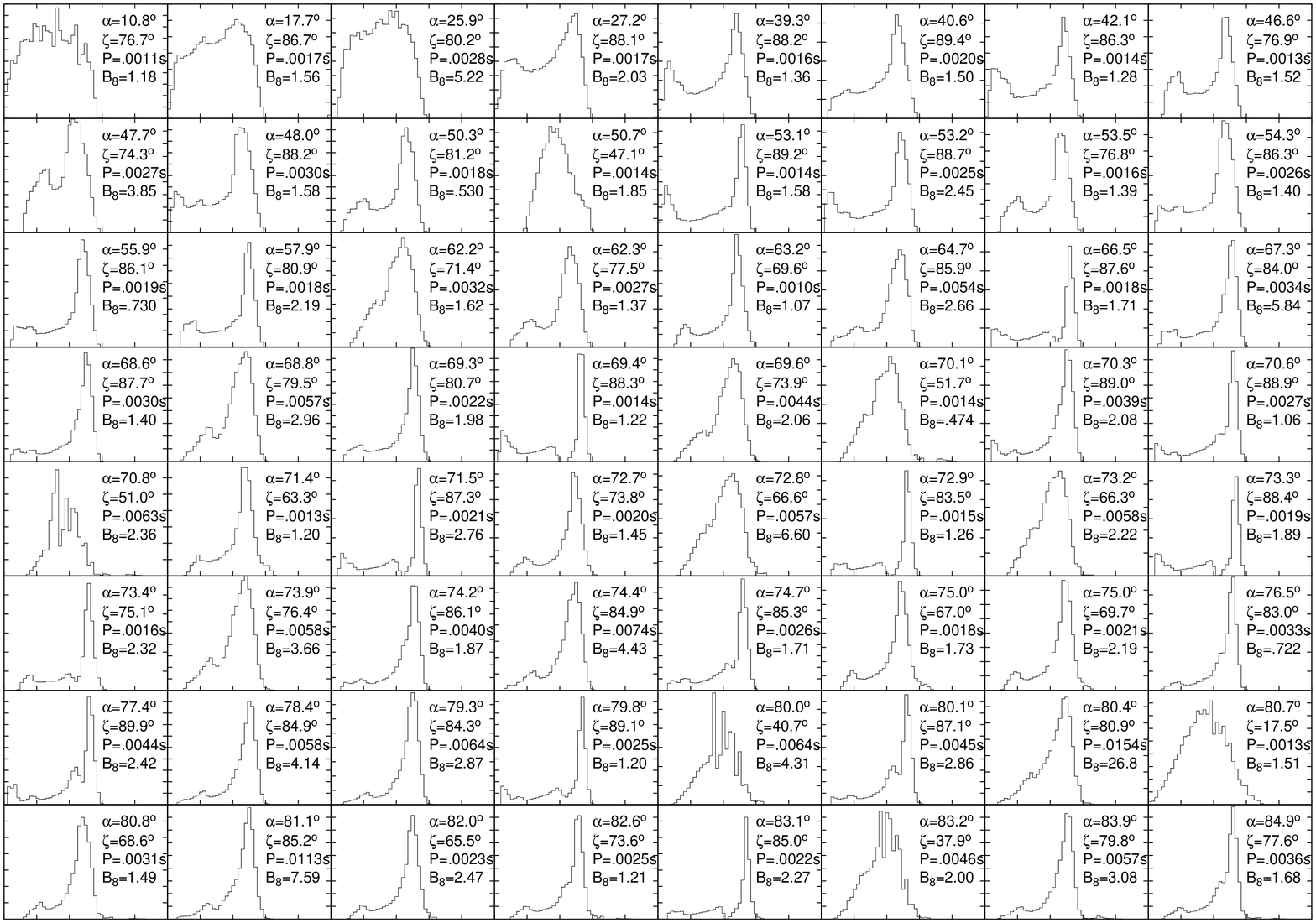}
\caption{The pulse profiles of 64 sample of the simulated $\gamma$-ray-
selected millisecond  pulsars.}
\label{pulsegMS}
\end{center}
\end{figure}

\label{lastpage}
\end{document}